\documentclass[letterpaper]{JHEP3}

\usepackage{epsf,epsfig}
\usepackage{amssymb}
\usepackage{graphicx}
\usepackage{amsmath}
\usepackage{amsfonts} 
 
\usepackage{graphicx}

\newcommand{\insertplot}[5]{\begin{figure}
 \hfill\hbox to 0.05in{\vbox to #5in{\vfill
 \inputplot{#1}{#4}{#5}}\hfill}
 \hfill\vspace{-.1in}
 \caption{#2}\label{#3}
 \end{figure}}
 \newcommand{\inputplot}[3]{
 \special{ps: plotfile #1}
}
\newcounter{fig}

\newcommand{\beq}{\begin{equation}}
\newcommand{\eeq}{\end{equation}}
\newcommand{\beqs}{\begin{eqnarray}}
\newcommand{\eeqs}{\end{eqnarray}}

\newcommand{\be}{\begin{equation}}
\newcommand{\ee}{\end{equation}}
\newcommand{\bea}{\begin{eqnarray}}
\newcommand{\eea}{\end{eqnarray}}

\newcommand{\la}{\lambda}
\newcommand{\del}{\delta}
\newcommand{\om}{\omega}
\newcommand{\ka}{\kappa}

\newcommand{\f}{\phi}
\newcommand{\vf}{\varphi}
\newcommand{\vr}{\varrho}

\newcommand{\al}{\alpha}

\newcommand{\si}{\sigma}

\newcommand{\pa}{\partial}

\newcommand{\vep}{\varepsilon}

\def\theequation{\arabic{equation}}

\newcommand{\re}[1]{(\ref{#1})}

\newcommand{\eins}{1\hspace{-0.56ex}{\rm I}}
\newcommand{\R}{\mathbb R}

\numberwithin{equation}{section}

\abstract{ 
We consider a class of generalizations of the Skyrme model to five spacetime dimensions ($d=5$), 
which is  defined in terms of an $O(5)$ sigma model. A special ansatz for the Skyrme field allows angular momentum to be present \textit{and} equations of motion with a radial dependence only. Using it, we obtain: 1) everywhere regular solutions describing localised energy lumps (\textit{Skyrmions}); 2) Self-gravitating, asymptotically flat, everywhere non-singular solitonic solutions (\textit{Skyrme stars}), upon minimally coupling the model to Einstein's gravity; 3) both static and spinning black holes with Skyrme hair, the latter with
rotation in two orthogonal planes, with both angular momenta of equal magnitude. In the absence of gravity we present an analytic solution that satisfies a BPS-type bound and explore numerically some of the non-BPS solutions. In the presence of gravity, we contrast the solutions to this model with solutions to a complex scalar field model, namely  boson stars and black holes with synchronised hair. 
Remarkably, even though the two models present key differences, and in particular the Skyrme model allows \textit{static} hairy black holes, when introducing rotation, the synchronisation condition becomes mandatory, providing further evidence for its generality in obtaining rotating hairy black holes. 
}

\preprint{ }

\title{   
Skyrmions, Skyrme stars and black holes with Skyrme hair in five spacetime dimension
} 
  
 \author{
{\large Yves Brihaye}$^{\dagger}$, 
{\large Carlos Herdeiro}$^{\ddagger}$,
{\large Eugen Radu}$^{\ddagger}$   
  and 
 {\large D. H. Tchrakian}$^{\star,  \diamond  }$
\\ 
\\
$^{\dagger}$
{\small Physique-Math\'ematique, Universite de
Mons-Hainaut, Mons, Belgium} 
\\
$^{\ddagger}$   
{\small  Departamento de F\'isica da Universidade de Aveiro and Centre for Research and Development in Mathematics and Applications (CIDMA), 
 Campus de Santiago, 3810-183 Aveiro, Portugal}
\\
$^{\star }${\small School of Theoretical Physics -- Dublin Institute for Advanced Studies, 10 Burlington
Road, Dublin 4, Ireland }  
\\
$^{\diamond}$ 
{\small  Department of Computer Science, Maynooth University, Maynooth, Ireland}

 }
 \begin{document}

\section{Introduction and motivation}

The Skyrme model has more than fifty years of history, starting from Skyrme's original construction 
and its basic solution, with unit baryon number \cite{Skyrme:1961vq,Skyrme:1962vh}.
It provided the very first explicit example of $solitons$ 
in a relativistic non-linear field theory in $d=3 + 1$ spacetime dimensions.
Such solutions, dubbed $Skyrmions$, have found interesting applications,
$e.g.$ as an effective description of low energy QCD \cite{Witten:1983tw}
and on the issue of proton decay \cite{Callan:1983nx}.

In its simplest version, the Skyrme model is described by a set of four scalars $\{\phi^a\}$,  
$a=1,..,4$, 
satisfying the sigma-model constraint
$ \phi^a \phi^a = 1$,  
with a target space\footnote{Usually the $SU(2)$ group element $U$ is employed, which in terms of the real field $\f^a$ is given by $U=\f^a\si_a$, with $\si_a=(i\si_i,\eins)$,
and $U^{-1}=U^{\dagger}=\f^a\tilde\si_a$, with $\tilde\si_a=(-i\si_i,\eins)$, where $\sigma_i$ are the standard Pauli matrices.} $S^3 \sim SU(2)$ and a Lagrangian density possessing a global $O(4)$ symmetry.
In addition to a standard ($quadratic$) kinetic term and a potential, the model contains 
an extra ($quartic$) term which is fourth order in derivatives\footnote{Higher derivative terms, up to the allowed $sextic$ kinetic term, can also be included as corrections~\cite{Floratos:2001bz,Floratos:2001ih}
and are generally expected,
see $e.g.$ the recent work 
\cite{Gudnason:2017opo}.}
allowing it to circumvent Derrick's no-go theorem for finite energy lump-like solutions in field theory~\cite{Derrick:1964ww}.  
These Skyrmions possess some topological properties, 
being  characterized by the homotopy class $\pi_3(SU(2))=\mathbb{Z}$.
Moreover, the energy functional of the Skyrme model has a Bogomol'nyi-type
bound,  in terms of the topological charge $B$ associated with the homotopy, which is 
identified as the baryon number.

The solutions of the Skyrme model have been studied intensively over the last decades.
In addition to spherical and axially symmetric configurations,
other classes of solutions have been identified, revealing sophisticated
geometrical structures with discrete symmetries only~
\cite{Braaten:1989rg,Hitchin:1995qw,Houghton:1995bs,Battye:1997qq,Battye:1997wf}.
Interestingly, for a topological charge $B>2$, the no-isometry configurations 
yield global  minima of the energy.
A detailed description of the flat spacetime Skyrmions can
be found in the monograph \cite{Manton:2004tk}.

\medskip

Skyrmions persist after taking their gravitational backreaction into account, within the Skyrme model minimally coupled to Einstein's gravity. The properties of such, hereafter \textit{Skyrme stars}, have been considered in~ 
\cite{Glendenning:1988qy,Droz:1991cx,Bizon:1992gb,Heusler:1991xx,Heusler:1993ci}.
Moreover, and following a generic rule in gravitating solitons
\cite{Kastor:1992qy,Herdeiro:2014ima,Herdeiro:2015waa}, 
these star-like configurations are compatible with the addition of a (small) horizon at their center. 
In the Skyrme model case,  this construction was carried out by Luckock and Moss~\cite{Luckock:1986tr}, and  provided the first physically relevant counter-example to the no-hair conjecture~\cite{Ruffini:1971bza}.
The construction in \cite{Luckock:1986tr} was performed in the probe limit, 
$i.e.$ a Skyrme test field on a Schwarzschild BH background, 
but subsequent work included the backreaction 
\cite{Luckock}.
This results in BHs with (primary) Skyrme hair, some of the solutions 
being stable against spherical linear perturbations
(a review of these aspects can be found in \cite{Volkov:1998cc}; 
 see also  
\cite{Adam:2016vzf,Dvali:2016mur,Gudnason:2016kuu,Gudnason:2015dca,Shnir:2015aba}
for more recent work).

Most of the known Skyrme-model solutions are static, but spinning generalizations have been constructed.
 Spinning flat spacetime Skyrmions were considered by Battye, Krusch and Sutcliffe \cite{Battye:2005nx}.
In contrast to the (conceptually simpler) spinning Q-ball solutions
\cite{Volkov:2002aj,Kleihaus:2005me},
for spinning Skyrmions the angular momentum $J$ is a continuous parameter that can be arbitrarily
small, so that they can rotate slowly and rotating solutions are continuously connected to static ones.
The effects of gravity on these spinning Skyrmions has been studied 
by Ioannidou,  Kleihaus and  Kunz
in \cite{Ioannidou:2006nn}, revealing, in particular, a number of configurations which do not have a flat space limit.
No rotating BHs with Skyrme hair, however, have been constructed so far, 
presumably due to the complexity 
of the numerical problem.

\medskip 

As we shall see,
the complexity just mentioned can be considerably alleviated 
by considering the generalisation of the Skyrme model to higher (odd) spacetime dimensions. 
Moreover, in recent years, the interest in field theory solutions
in $d \neq 4$ increased significantly.  
A recurrent lesson has been that well known results in $d = 4$ physics
do not have a simple extension to higher dimensions.
For example, the BH solutions in $d\geqslant 5$ models of gravity are less constrained,
with a variety of allowed horizon topologies \cite{Emparan:2008eg}.
In the Skyrme case, however, the only  other dimensions considered so far in the literature is $d=3$~
\cite{Piette:1994jt,Piette:1994ug,Piette:1994mh,Arthur:1996ia}, even though (gravitating) Skyrmions should also 
have $d\geqslant 5$ generalizations.

In this paper we shall consider a higher dimensional Skyrme model with the main goal 
of testing the existence of rotating BHs with Skyrme hair, but also to understand how the spacetime dimension affects standard results for Skyrme physics. 
The Skyrme system  to be addressed 
is a generalisation  of the usual $d=4$ Skyrme model,
containing higher derivative terms in  addition to the standard ones.
This can be done, in principle, in all (including even) dimensions; for concreteness we restrict our attention to the $d=4+1$ case, 
where we have carried out a detailed numerical study. 
A technical advantage of this case is the possibility to consider 
configurations with two equal angular momenta,  for which there is a symmetry enhancement.
As such, the problem results in a system of ordinary differential equations
(ODEs) which are easier to study.
 
 We shall also contrast the Skyrme model with the better known and conceptually simpler model of a complex scalar field, for which one also finds flat spacetime field theory solutions ($Q$-balls~\cite{Coleman:1985ki,Volkov:2002aj,Kleihaus:2005me}), gravitating solitons (\textit{boson stars}~\cite{Schunck:2003kk}) and 
hairy BHs (Kerr BHs with scalar hair~\cite{Herdeiro:2014goa,Herdeiro:2015gia,Herdeiro:2015tia}). In particular for the latter there are no static hairy BHs. But in agreement with the latter, rotating BHs with Skyrme hair also require a \textit{synchronisation} condition, as described below.

\medskip
 The paper is organized as follows.
In Section 2 we introduce the model together with a codimension-1
Skyrme field ansatz. The non-gravitating and gravitating solitons of the model -- Skyrmions and  Skyrme stars -- are discussed in Sections 3 and 4.
 BHs with Skyrme hair are presented in Section 5. 
In Sections 3-5 we study both static and rotating solutions.
We give our conclusions and remarks in Section 6.
Numerical solutions with a quartic term in the Skyrme action are reported in 
Appendix A.
Appendix B contains
a discussion of the Skyrme model for a general number of spacetime dimensions.

\medskip
 {\bf Conventions and numerical method.}
%
Throughout the paper, mid alphabet latin letters $i,j,\dots$ 
label spacetime coordinates, 
  running from $1$ to $5$ (with $x^5=t$);
Early latin letters, $a,b,\dots $ label the internal indices of the scalar field multiplet.
As standard, we use Einstein's summation convention, but to alleviate  notation, no distinction is made between covariant and contravariant \textit{internal} indices.

The background of the theory is Minkowski spacetime, where the spatial $\mathbb{R}^4$ is written in terms of bi-polar spherical coordinates,
\begin{eqnarray}
ds^2=dr^2+r^2(d\theta^2+\sin^2\theta d\varphi_1^2+\cos^2\theta d\varphi_2^2)-dt^2~,
\end{eqnarray}
where $\theta  \in [0,\pi/2]$ is a polar angle interpolating between the two orthogonal 2-planes and $(\varphi_1,\varphi_2) \in [0,2\pi]$ are azimuthal coordinates, one in each 2-plane. $r$ and $t$ denote the radial and time coordinate, $0\leqslant r<\infty$ and $-\infty <t<\infty$, respectively.

For most of the solutions the numerical integration was carried out using a standard shooting method.
In this approach, we evaluate the
initial conditions at $r = 10^{-5}$ (or $r=r_H+10^{-5}$) for global tolerance $10^{-14}$, adjusting for shooting parameters
and integrating towards $r\to \infty$.
The spinning gravitating solutions were found by using a professional software package \cite{COLSYS}. 
This solver employs a collocation method for boundary-value ODEs 
and a damped Newton method of quasi-linearization. 
A linearized problem is solved
at each iteration step, by using a spline collocation at Gaussian points.
An adaptive mesh selection procedure is also used, 
such that the equations are solved on a sequence of meshes until the
successful stopping criterion is reached.


 \section{The $O(5)$ Skyrme model in $d=4+1$ dimensions}

The Skyrme model can be generalised to an arbitrary number of dimension -- $cf.$ Appendix B.
Restricting to
 $d=4+1$ spacetime dimensions, the Skyrme model
is defined in terms of the $O(5)$ sigma model real fields $\{\phi^a\}$, $a=1,\dots,5$, satisfying
the constraint $\sum_a\phi^a\phi^a = 1$.  
It proves useful to introduce the following notation for all the allowed kinetic terms, quadratic, quartic, sextic and octic:
\begin{eqnarray}
\nonumber
&&
\f(1)=\phi^a_{i}\equiv \partial_i \phi^a\ ,
\\
\nonumber
&&
\f(2)=\phi^{ab}_{i j}\equiv  \phi^a_{i} \phi^b_{j}-\phi^a_{j} \phi^b_{i}\ ,
\\
&&
\nonumber
\f(3)=\phi^{abc}_{i j k}\equiv \phi^{ab}_{i j} \phi^c_{k}
                  +\phi^{ab}_{jk} \phi^c_{i}
                  +\phi^{ab}_{ki} \phi^c_{j} \ ,\\
&&
\nonumber
\f(4)=\phi^{abcd}_{i j kl}\equiv \phi^{abc}_{i jk} \phi^d_{l}
                  +\phi^{abc}_{jkl} \phi^d_{i}
                  +\phi^{abc}_{kli} \phi^d_{j} \ ,
\end{eqnarray}
in terms of which the kinetic terms are written:
\begin{eqnarray}
&&
\nonumber
{\cal F}^2\equiv \phi^a_{i_1} \phi^{a}_{i_2} g^{i_1 i_2}\ ,
\\
\label{calF}
&&
{\cal F}^4\equiv \phi^{ab}_{i_1 j_1} \phi^{ab}_{i_2 j_2} g^{i_1 i_2}g^{j_1 j_2} \ ,
\\
&&
\nonumber
{\cal F}^6 \equiv \phi^{abc}_{i_1 j_1 k_1} \phi^{abc}_{i_2 j_2 k_2} g^{i_1 i_2} g^{j_1 j_2}g^{k_1 k_2} \ ,
\\
&&
\nonumber
{\cal F}^8 \equiv \phi^{abcd}_{i_1 j_1 k_1l_1} \phi^{abcd}_{i_2 j_2 k_2l_2} g^{i_1 i_2} g^{j_1 j_2}g^{k_1 k_2}g^{l_1 l_2} \ .
\end{eqnarray}
Observe that  
$\phi^{a,i}=\phi^{a}_j g^{ij}$,
$\phi^{ab,ij}=\phi^{ab}_{kl}g^{ik}g^{jl}$, 
$etc.$, where $g_{ij}$ is the metric tensor of the five dimensional background geometry.

We shall always include in the action the quadratic term ${\cal F}^2$. 
Then,
for most solutions in this paper, as will be justified in Section~\ref{secsca} 
by a virial identity, only the sextic term ${\cal F}^6$ is also needed, so we will eschew the octic terms, 
which do not bring any qualitative features to the solutions.
We shall likewise drop the quartic term ${\cal F}^4$, though in Appendix {\bf A} we verify that its inclusion does not change the general features.
The sole exception to this pattern is the BPS solution of Section~\ref{bpssol} which relies only on the quartic term.

The Lagrangian density of the model considered throughout is
\begin{eqnarray}
\label{LS}
{\cal L}_S= \frac{\lambda_1}{2}{\cal F}^2+\frac{\lambda_2 }{4}{\cal F}^4+\frac{\lambda_3}{36} {\cal F}^6+\lambda_0 V (\phi^a)\ ,
\end{eqnarray}
where $V$ is the Skyrme potential whose explicit form will be discussed later and $\la_i\geqslant 0$ are coupling constants. Observe these are dimensionful constants: 
$[\lambda_0] = {\rm length}^{-5}$,
$[\lambda_1]= {\rm length}^{-3}$,
$[\lambda_2]= {\rm length}^{-1}$
and
$[\lambda_3]=  {\rm length}$. 
Inclusion of the 
 potential term is mandatory for rotating solutions.  
%

From the Lagrangian (\ref{LS}) it follows that the scalars $\phi^a$ satisfy the
 Euler-Lagrange equations
\be
\label{EL1}
\left(\del^{d a}-\f^{d }\f^{a}\right)
\left\{
2\la_1\,\nabla^{i}\f_{i}^{a}
+8\la_2\,\f_{i}^{b,k}\nabla^{j}\f_{jk}^{ab, i}
+9\la_3\,\f^{bc,jk}\nabla^{i}\f_{ijk}^{abc}
+\la_0\,\frac{\pa V}{\pa\f^{a}}
\right\}=0 \ .
\ee
%
\subsection{A BPS bound and the topological charge}
\label{bpssol}

The model (\ref{LS}) with $\lambda_0=\lambda_1=\lambda_3=0$, $i.e.$ the action
\begin{eqnarray}
\label{actionSd}
I_S=\frac{\lambda_2}{4} \int d^5 x \sqrt{-g}
\phi^{ab}_{i_1 j_1} \phi^{ab}_{i_2 j_2} g^{i_1 i_2}g^{j_1 j_2}~,
\end{eqnarray}
possesses some special properties, provided the
spacetime  geometry is ultra-static ($g_{tt}=-1$) and the Skyrme ansatz has no time dependence.
Then the resulting system lives effectively on a four dimensional 
space with Euclidean signature 
(which, however, can be curved),
being conformal invariant.
 
After defining the
two-form Hodge dual of $\phi^{ab}_{i j }$ as
\begin{eqnarray}
\label{sd1}
^{\star} \phi^{ab}_{i j }=\sqrt{-g}\epsilon^{ab a_1 b_1 c}\epsilon_{iji_1j_15} \phi^{a_1b_1, i_1  j_1 }\phi^c~,
\end{eqnarray}
we can state the Bogomol'nyi inequality
\be
\label{Bogi4}
\left|\f_{i_1i_2 }^{a_1a_2 }\mp \, ^{\star} \phi_{i_1i_2 }^{a_1a_2 }\right|^2\geqslant 0\ ,\quad  
\ee
which implies that the mass-energy of the model
is bounded from bellow
\be
\label{bound1}
M\geqslant  \frac{\lambda_2}{4}   B \ ,
\ee
where $B$ is the topological charge
\be
\label{top-charge}
 B=\int d^4 x \sqrt{-g}~\rho_T,~~{\rm with}~~~\rho_T=\frac{1}{64\pi^2} \f_{i_1i_2 }^{a_1a_2 }~  ^{\star}\f^{a_1a_2,i_1i_2  },
\ee
 $\rho_T$
being
the topological charge density. 
Also, the total mass-energy of $d=5$ solutions is defined as 
\be
M=\frac{\lambda_2}{4} \int d^4 x \sqrt{-g}~{\cal F}^4 \ ,
\ee
while the topological current is
 \be
B^k=\frac{1}{\sqrt{-g}}
\frac{1}{64\pi^2} 
\epsilon^{iji_1j_1 k} 
 \epsilon_{ab a_1 b_1 c} 
\f_{i j }^{ab} \phi^{a_1b_1}_ {i_1  j_1 }\phi^c,
\ee
(with $B^t\equiv \rho_T$).

As we shall see in the next Section, self-dual solutions saturating the above bound 
exist\footnote{This contrasts with the Skyrmions on $\R^3$ where no BPS solitons exist, while they do  on $S^3$~\cite{Jackson:1988bd}.},
being solutions of the 1st order equations
\begin{eqnarray}
\label{sd0}
 \phi^{ab}_{i  j }=\pm ^{\star} \phi^{ab}_{i  j }~.
\end{eqnarray}
It is clear  that the bound (\ref{bound1})
holds as well for the general Lagrangian (\ref{LS}),
in which case it can never be saturated, 
since the contribution of the supplementary terms is always positive.

 \subsection{Coupling to gravity}
 
The action of the $d=5$
Einstein-Skyrme model reads
\begin{eqnarray}
\label{action}
I=\int d^5 x \sqrt{-g}\left(\frac{R}{16 \pi G}-{\cal L}_S \right),
\end{eqnarray}
where ${\cal L}_S$ is the Lagrangian density (\ref{LS}) for the Skyrme sector
and $G$ is Newton's constant.

Variation of (\ref{action}) w.r.t. the metric tensor leads to the Einstein equations 
\begin{eqnarray}
\label{grav-eq}
R_{ij}-\frac{1}{2}g_{ij}=8 \pi G~T_{ij},
\end{eqnarray}
where the energy-momemtum tensor is
\begin{eqnarray}
 T_{ij} = \lambda_0 T_{ij}^{(0)}+\lambda_1 T_{ij}^{(1)}+\lambda_2 T_{ij}^{(2)}+\lambda_3 T_{ij}^{(3)} \ ,
\end{eqnarray}
in terms of the contributions of the distinct terms in (\ref{LS}), which read
\begin{eqnarray}
\nonumber
&&
 T_{ij}^{(0)} =-g_{ij} V (\phi^a),
\\
\nonumber
&&
 T_{ij}^{(1)} = \phi^a_{i} \phi^a_{j}-\frac{1}{2}g_{ij}{\cal F}^2,
\\
\label{tij}
&&
 T_{ij}^{(2)} =   \phi^{ab}_{i k_1} \phi^{ab}_{j k_2}g^{k_1 k_2}-\frac{1}{4}g_{ij}{\cal F}^4 ,
\\
\nonumber
&&
 T_{ij}^{(3)} = \frac{1}{6}\left(\phi^{abc}_{i k_1 l_1} \phi^{abc}_{j k_2 l_2}g^{k_1 k_2} g^{l_1 l_2}-\frac{1}{6}g_{ij}{\cal F}^6 \right).
\end{eqnarray}
 As usual, for a given ansatz, the gravity equations (\ref{grav-eq}) are solved together with the matter equations (\ref{EL1}), subject to some physical requirements ($e.g.$asymptotic flatness and finiteness of the total mass).

\subsection{A codimension-1 Skyrme field ansatz}

The $O(5)$ 
solutions in this work are constructed within a Skyrme fields ansatz 
in terms of a single function $F(r)$:
\begin{eqnarray}
\nonumber
&&
 \phi^1+i\phi^2=\sin F(r)\sin\theta \, e^{i(\vf_1-\om t)}\ ,
\\
&&
\label{phi}
 \phi^3+i\phi^4=\sin F(r)\cos\theta \, e^{i(\vf_2-\om t)}\ ,
\\
\nonumber
&&
 \phi^5=\cos F(r)  \ .
\end{eqnarray}
Here, $w\geqslant 0$ is an input parameter -- the frequency of the fields.
The corresponding expression of the 
 topological charge density is:
\begin{eqnarray}
\label{top-sph}
\rho_T =  \frac{1}{16}\frac{d}{dr}
\left [
\cos(3F)-9\cos F
\right ]~.
\end{eqnarray}

For the Skyrme potential we shall take the 
usual 'pion mass'-type  
\begin{eqnarray}
V=  1-\phi^5 =2\sin^2\left[\frac{F(r)}{2}\right] \ ,
\end{eqnarray}
which is a natural generalization of that used in the  $d=4$ model.

A remarkable feature of the ansatz (\ref{phi}), first suggested in~\cite{Hartmann:2010pm}, 
albeit for  a complex doublet rather than a Skyrme field,
 is that for any geometry in this work
the angular dependence is factorized in a consistent way,
and the Skyrme equations (\ref{EL1})
reduce to a single ODE for the function $F(r)$.

\section{Flat spacetime Skyrmions }

We start by considering solutions in the probe limit, 
$i.e.$ we solve the Skyrme equations on a fixed spacetime background.
Apart from being technically simpler,
we shall find that they possess already a number of basic properties
of the corresponding gravitating generalizations.

\subsection{A spherical BPS Skyrmion in the quartic model}

In the simplest spherically symmetric case
and a flat background,
the model  with only a quartic term~(\ref{actionSd}) allows a simple analytical solution,
found with the ansatz (\ref{phi}) (with $\om=0$). Then 
the first order eqs. (\ref{Bogi4}) reduce to\footnote{It is interesting to note that the solution (\ref{ex1}) is related to the radially symmetric BPST instanton~\cite{Belavin:1975fg} described by the form factor $w(r)$,
$via$ $w(r)=\cos F$.
}
\begin{eqnarray}
\label{ex12}
F' \pm \frac{\sin F}{r}=0 \ .
\end{eqnarray}
Restricting to the plus sign\footnote{The minus sign solution reads $F(r)=\pi-2\arctan ({r_0}/{r})$
and possesses similar properties.}, the solution  of the above equation reads 
\begin{eqnarray}
\label{ex1}
F(r)= 2\arctan \left(\frac{r_0}{r}\right)~,
\end{eqnarray}
with $r_0>0$ an arbitrary parameter.
This is an everywhere regular configuration, with $F(r)$ interpolating between $\pi$ (at $r=0$) and zero (at $r=\infty$).  
Its energy density  is
\begin{eqnarray}
\label{ex2}
\rho(r)=-T_t^t= \frac{96 \lambda_2 r_0^4}{(r^2+r_0^2)^4}\ ,
\end{eqnarray}
while all other components of $T_i^j$ vanish.
The  total mass of solution is
\begin{eqnarray}
\label{ex3}
M=\frac{\lambda_2}{4} ~.
\end{eqnarray}
It would be interesting to investigate the existence of static
non-selfdual solutions of the simple model (\ref{actionSd}).
They are expected to exist,
in analogy with  
instanton--anti-instanton solutions to  Yang-Mills theory \cite{Radu:2006gg}.

\subsection{The general model }
%

\subsubsection{The effective action and densities}
 
The simple quartic model (\ref{actionSd}) is too restrictive.
Indeed, the self-duality eqs. (\ref{sd0}) cannot be satisfied for $g_{tt}\neq -1$ or $\om\neq 0$.
 The situation in this restrictive case is the same as what occurs in the Einstein-Yang-Mills system 
\cite{Volkov:2001tb}, 
where the solutions of the usual Yang-Mills (YM) model $F(2)^2$
do not survive when considering their backreaction
on the spacetime geometry. 
This is because the scaling requirement is violated. 
In that case, this defect was remedied by adding the higher-order YM term $F(4)^2$
and regular gravitating YM solutions were constructed
for $d=6,7$ spacetime dimension  \cite{Brihaye:2002hr}. 
Subsequently, the case $d=5$ was considered as well \cite{Brihaye:2002jg}, 
where also BH solutions were constructed.

For the Skyrme system,
the ${\cal F}^4$ density scales as $L^{-4}$, while the usual gravity scales as $L^{-2}$. 
Thus, as in the YM case, it is necessary to add higher-order terms.
%
%
 %
In the general case with all $\lambda_i\neq 0$  
and a Skyrme ansatz given by (\ref{phi}), 
one can show that the equation for $F(r)$
can also be derived from the effective action 
\be
\label{Seff}
S =\int_0^{\infty} dr L_{eff} \ , \qquad {\rm with}  \qquad
L_{eff}= r^3
\left[ \lambda_0(1-\cos F) +\frac{\lambda_1}{2} \bar {\cal F}^2
+\frac{\lambda_2}{4} \bar{\cal F}^4
+\frac{\lambda_3}{36} \bar{\cal F}^6
\right] \ ,
\ee
where
\begin{eqnarray}
&&
\nonumber
\bar{\cal F}^2=F'^2+\left(\frac{3}{r^2}-\om^2\right)\sin^2F\ ,
\\
&&
\bar{\cal F}^4=\left[ \left(\frac{3}{r^2}-\om^2\right)
F'^2
+\left(\frac{3}{r^2}-2\om^2\right)\frac{\sin^2F}{r^2}
\right] 4\sin^2F\ ,
\\
&&
\nonumber
\bar{\cal F}^6=\left[ \left(\frac{3}{r^2}-2\om^2\right)
F'^2
+\left(\frac{1}{r^2}- \om^2\right) \frac{\sin^2F}{r^2}
\right]\frac{ 36\sin^4F}{r^2} \ .
\end{eqnarray}
Despite the fact that this describes 
an
effective one-dimensional system,
the configurations with $\omega \neq 0$
are $not$ spherically symmetric\footnote{This contrasts with the complex scalar fields model in \cite{Hartmann:2010pm},
which possesses spherically symmetric Q-ball solutions
supported by the harmonic time dependence of the fields.} and carry an
angular momentum density, $j$, 
\begin{equation}
\label{j-dens}
\small
j\equiv \frac{T_{\varphi_1}^t }{\sin^2\theta}  = \frac{T_{\varphi_2}^t}{\cos^2\theta} =
\omega \sin^2 F
\bigg[
\lambda_1+2\lambda_2\left(F'^2+  \frac{2 \sin^2F}{r^2} \right)
+2\lambda_3  \frac{ \sin^2F}{r^2}\left(2 F'^2+ \frac{\sin^2F}{r^2} \right)
\bigg].
~~{~~~}
\end{equation}
Their energy density, $\rho$, is, however, spherically symmetric, with  
\begin{eqnarray}
\nonumber
\rho=-T_{t}^t&=&
 \lambda_0(1-\cos F) +\frac{1}{2}\lambda_1\left[F'^2+ \left(\frac{3}{r^2}+\om^2 \right)\sin^2F \right]
\\
\label{ro-dens}
&&
+\frac{1}{4}\lambda_2  \left[  \left(\frac{3}{r^2}-\om^2 \right)
F'^2
+ \left(\frac{3}{r^2}+2\om^2 \right)\frac{\sin^2F}{r^2}
\right] 4\sin^2F 
\\
&&
\nonumber
+\frac{1}{36}\lambda_3 
\left[  \left(\frac{3}{r^2}+2\om^2 \right)
F'^2
+\left(\frac{1}{r^2} + \om^2\right)\sin^2F
\right]\frac{ 36\sin^4F}{r^2} \ .
\end{eqnarray}

As usual in the probe limit, the total mass, $M$, and angular momenta, $J$, of a soliton are defined as 
\begin{equation}
\label{MJ}
\small
M=-\int d^4 x \sqrt{-g}T_t^t=2\pi^2 \int_0^\infty dr~r^3 \rho,~~
J_1=J_2=J= \int d^4 x \sqrt{-g}T_{\vf_{1,2}}^t=\pi^2 \int_0^\infty dr~r^3 j.  
\end{equation}
Following $d=5$ BH physics conventions,
we define also the reduced angular momentum of a spinning configuration as
\begin{eqnarray}
\label{red-j}
 \bar j\equiv \frac{27\pi}{8}\frac{J^2}{M^3}~.
\end{eqnarray}

\subsubsection{A virial identity and scaling}
\label{secsca}

The form (\ref{Seff}) of the reduced allows the derivation of an useful Derrick-type virial relation.
Let us assume the
existence of a globally regular solution $F(r)$, with suitable boundary
conditions at the origin and at infinity. 
Then each member of the 1-parameter family
$F_{\lambda}(r)\equiv F(\lambda r)$
assumes the same boundary values at
$r=0$ and $r=\infty$
and  the action $S_\lambda\equiv S(F_\lambda)$
must have a
critical point at $\lambda=1$,
$i.e.$
$
[
dS_{\lambda}/d\lambda
]_{\lambda=1}=0
$.
This results in the following virial identity satisfied by the
finite energy solutions of the field equations
\begin{eqnarray}
\label{virial1}
&&
\int_0^\infty
dr~r^3
\left
[
\lambda_1  \left ( F'^2+  \frac{3  \sin^2F}{r^2}  \right)
+4\lambda_0 (1-\cos F)
\right ]
=
\\
\nonumber
&&
\int_0^\infty
dr~r^3
\left\{
2\lambda_3 \frac{ \sin^4F}{r^4}
\left( 
3F'^2
+ \frac{\sin^2F }{r^2}  
\right)
+2\omega^2 \sin^2 F \left[
\lambda_1 
+2\lambda_2 \left( F'^2+  \omega^2 \frac{ \sin^2F}{r^2}\right)
\right]
\right\}~.
\end{eqnarray}
The positivity of all terms in the previous relation shows that the existence of $d=5$  Skyrmions can be attributed to a balance 
between: the attractive interaction provided by the 
usual kinetic term in the  Skyrme  action together with the potential (left hand side terms); 
and a repulsive interaction provided by the sextic term, plus the centrifugal force in the rotating case (right hand side terms).
Also, one can see that, as anticipated above, the contribution of the quadratic term occurs with an overall $w^2$-factor only, 
and is not mandatory for the existence of solutions.

\medskip

The Skyrme Lagrangian (\ref{LS}) 
contains four input parameters $\lambda_i$.
However, the constant multiplying the 
quadratic term can be taken as an overall  
factor for the Skyrme action.
Also, the equation for $F$
is invariant under the transformation 
$r\to \tau r$,
$\lambda_0/\lambda_1 \to  \tau^2 \lambda_0/ \lambda_1$,
$\lambda_2/\lambda_1 \to \lambda_2/ (\tau^2 \lambda_1)$,
$\lambda_3/\lambda_1 \to \lambda_3/ (\tau^4 \lambda_1)$
together with
$w \to w/\tau$,
which can be used to
 fix the value of one of the constants $\lambda_0,\lambda_2$ or $\lambda_3$.
Then
following the $d=4$ case, we define a characteristic length $L$ and mass $M_S$ of the Skyrmion system
as given by the
the constants multiplying the quadratic and potential terms,
with
\begin{eqnarray}
L \equiv \sqrt{\frac{\lambda_1}{\lambda_0}} \ , \qquad M_S \equiv \frac{\lambda_1^2}{\lambda_0}~,
\end{eqnarray}
the numerical results being obtained in units set by $L$ and $M_S$.
However, to avoid cluttering the output with a complicated dependence
of $L,M_S$
we shall ignore these factors in the displayed numerical results.

The problem still contains 
two free constants which multiply the quartic and sextic terms in the Skyrme action.
Moreover, in the presence of gravity,  one extra parameter occurs.
Therefore, the determination of the domain of existence of the solutions would be a lengthy task. 
In this work, in order to
 to simplify the picture,
we have chosen
to solve a model without the quartic term and with a unit value for the parameter multiplying the sextic term.
Thus, all reported numerical results below are found for the following choice of the coupling constants
\begin{eqnarray}
\lambda_0= \lambda_1=\lambda_3=1,~~\lambda_2=0\ .
\end{eqnarray} 
To provide evidence that this choice does not restrict the generality of our results, we present, in Appendix A, solutions including the presence of a quartic term
in the action, $\lambda_2\neq 0$, which indeed does not appear to affect the qualitative properties of the solutions.

\subsubsection{The solutions}

The function $F(r)$ satisfies a non-enlightening second order 
differential equation which we shall not display here
(its $\om=0$ limit can be read off by setting $N(r)=\sigma(r)=1$ in (\ref{eqF})).
This equation does not seem to possess exact solutions and it is solved numerically
with the boundary conditions\footnote{Solutions  
with $F(r)$ interpolating between $k \pi$ (with $k>1$) 
at $r=0$ and $F(\infty)=0$ 
do also exist. However, they are more massive and likely to be unstable.
} 
$F(0)=\pi$ and $F(\infty)=0$,
which follow from finite energy 
and regularity
requirements.

\vspace*{-0.1cm}
 {\small \hspace*{3.cm}{\it  } }
\begin{figure}[ht]
\hbox to\linewidth{\hss%
	\resizebox{8cm}{6cm}{\includegraphics{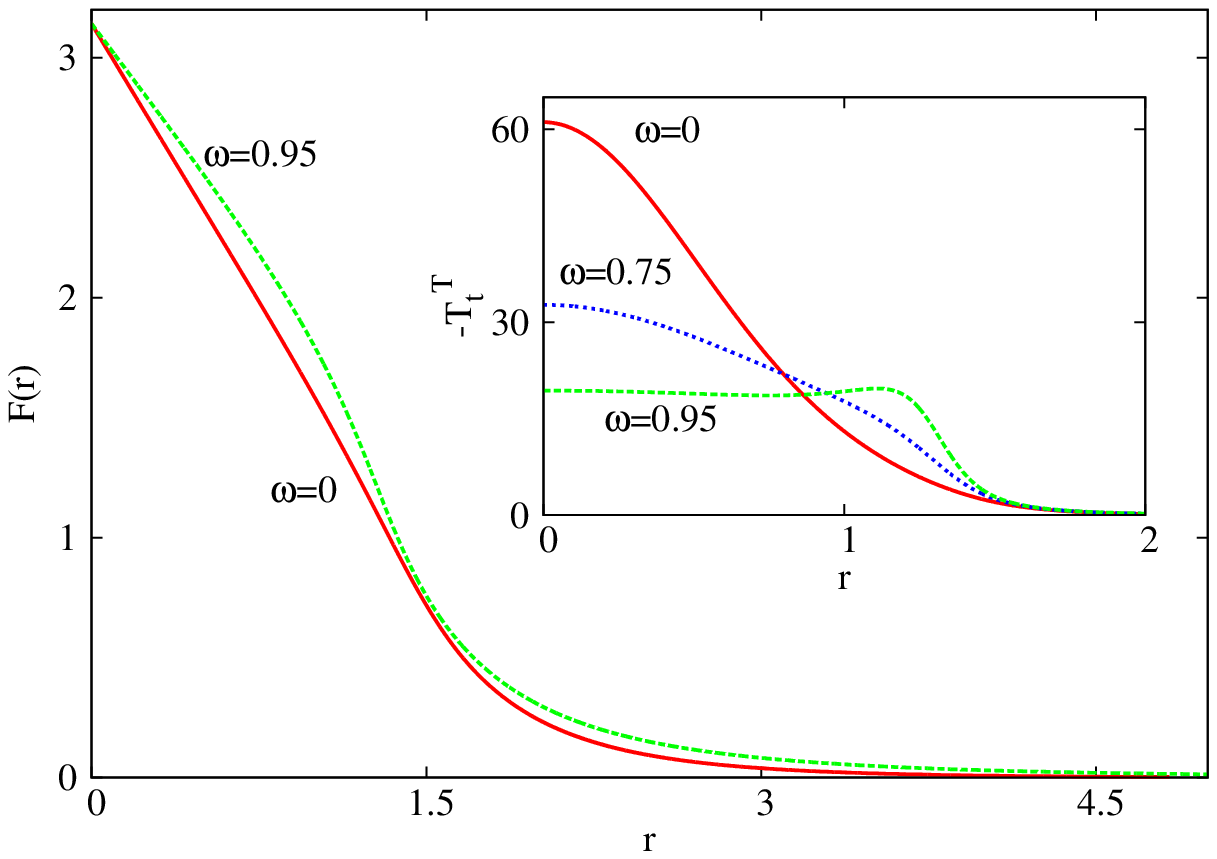}}
\hspace{10mm}%
        \resizebox{8cm}{6cm}{\includegraphics{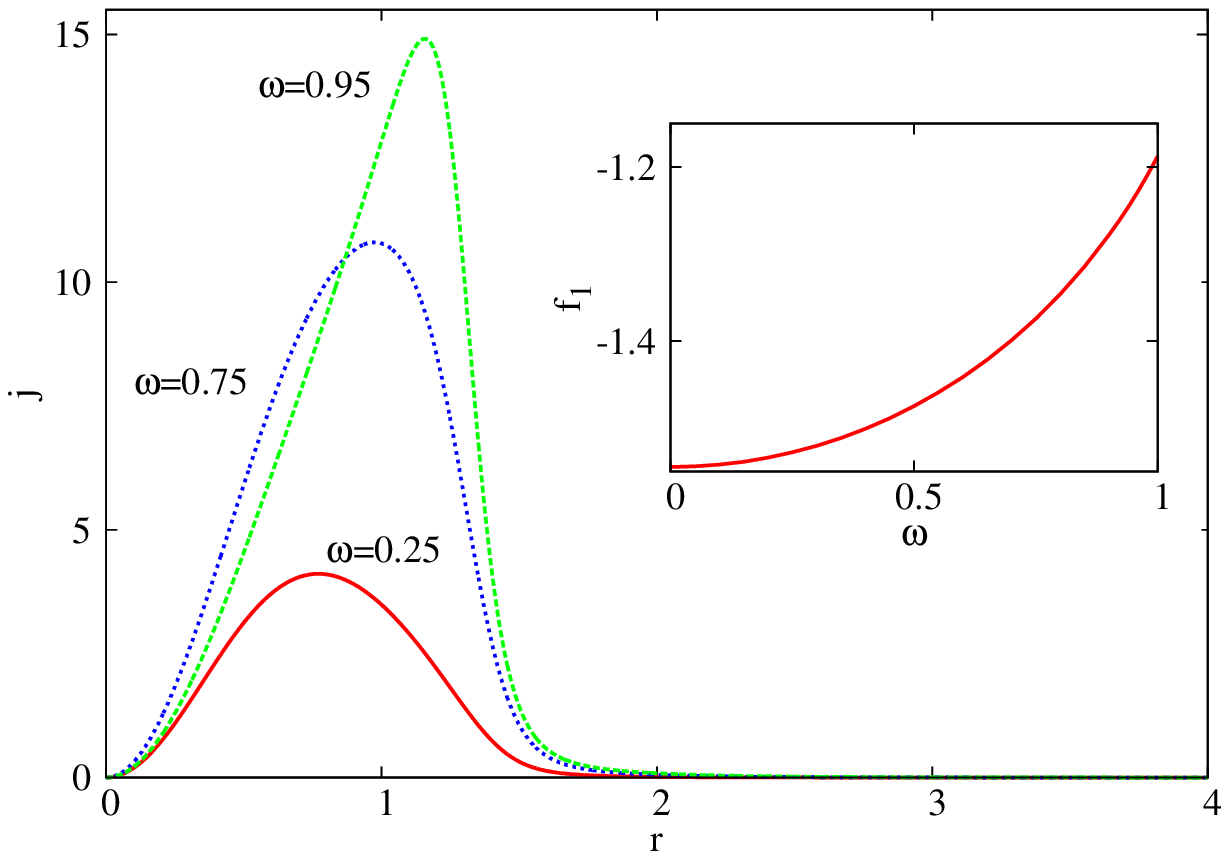}}	
\hss}
 \caption{\small
{\it Left panel:} The  profiles of $d=5$ static and spinning Skyrmions, on a flat spacetime background
are shown together with their energy density.
{\it Right panel:} The  distribution of the angular momentum density for spinning  Skyrmions. 
The inset shows the values of the parameter $f_1$ 
which enters the small$-r$ expression  of the solutions. 
  }
\end{figure}

The asymptotics of $F(r)$ can be systematically constructed in both regions, near the
origin and for large $r$.
The  corresponding expression for small $r$ is
\be
\label{orS}
F(r)=\pi+f_1 r+f_3 r^3+\mathcal{O}(r^5),~~{\rm with}~~
f_3=-f_1\frac{\lambda_0+\lambda_1(2f_1^2+\om^2)+6f_1^4(\lambda_2-\lambda_3 \om^2)}{12[\lambda_1+6f_1^2(\lambda_2+f_1^2\lambda_3)]} \ ,
~~{~~}
\ee
in terms of a single undetermined parameter $f_1<0$.
One notice that the energy density at $r=0$ is nonzero, with
\begin{eqnarray}
\rho(0)=2(\lambda_0+f_1^2\lambda_1+3f_1^4\lambda_2+2f_1^4 \lambda_3)~,
\end{eqnarray}
while the angular momentum density $j$, given by (\ref{j-dens}), vanishes as $\mathcal{O}(r^2)$.

An approximate solution valid for large $r$ 
can be written in terms of the modified Bessel function of the second kind, $K$,
\begin{eqnarray}
\label{infS}
F(r)\sim \frac{c}{r}K \left[2,r\sqrt{\frac{\lambda_0}{\lambda_1}-\om^2}  \right] \to c\sqrt{\frac{\pi}{2}}\frac{ e^{-r\sqrt{\frac{\lambda_0}{\lambda_1}-\om^2}}}{r^{3/2}}+\dots,
\end{eqnarray}
(with $c$ a constant),
which shows the existence of an upper bound on the scalar field
frequency, 
$\om\leq \sqrt{\lambda_0/\lambda_1}=1/L$,
the solutions becoming delocalized for larger values of $\om$.
Therefore, similar to other examples of spinning scalar solitons
($e.g.$ \cite{Volkov:2002aj,Kleihaus:2005me}),
the presence of a potential term in the action, $\lambda_0\neq 0$,
is a pre-requisite for the existence of finite mass solutions.
Note, however, that we have found numerical
evidence for the existence of static solitons with $V=0$, 
which decay as $1/r^3$ at infinity. 
 
Solutions interpolating 
between
(\ref{orS})
and
(\ref{infS})
are easily constructed -- Figure 1.
In our approach, the control ``shooting" parameter 
is $f_1$
which enters the near origin expansion (\ref{orS}) -- see Figure 1, inset of right panel.
For a given frequency $\om$,
a single nodeless solution is found  for a special value of $f_1$.
The profiles of the energy density and of the angular momentum density of typical solutions are shown in Figure 1, including the static one, which has  $\om=0$.
In Figure 2 
we display the total mass and angular momentum of the spinning Skyrmions as a function of their frequency.
One can see that both quantities increase monotonically with $\om$, with a smooth
$\om\to 1$ limit which maximizes the values of mass and angular momentum.
Also, at least for the considered values of the coupling constants,
the Skyrmions are never fast spinning objects,
with  a reduced angular momentum $\bar j$ always much smaller than one.

{\small \hspace*{3.cm}{\it  } }
\begin{figure}[t!]
\hbox to\linewidth{\hss%
	\resizebox{8cm}{6cm}{\includegraphics{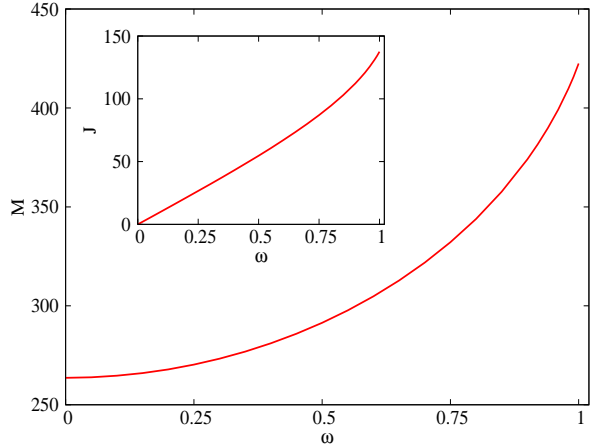}}
\hss}
\caption{\small 
The total mass, $M$, and total angular momentum, $J$, of static ($\omega=0$) and spinning Skyrmions on a flat spacetime background are shown 
as a function of their frequency $\omega$.}
\end{figure}

\section{Skyrme stars}

\subsection{Spherical stars}

The above solutions possess gravitating generalizations,
which are found by solving the Skyrme equation (\ref{EL1})
together with the Einstein equations (\ref{grav-eq}). 
A suitable metric for spherically symmetric configurations
reads
\be
\label{sph-metric}
ds^2=\frac{dr^2}{N(r)}+r^2(d\theta^2+\sin^2\theta d\varphi_1^2+\cos^2\theta d\varphi_2^2)-N(r)\sigma^2(r)dt^2~,
~{\rm where}~N(r)\equiv 1-\frac{m(r)}{r^2},~~~~{~~~~}
\ee
the function $m(r)$ being related to the local mass-energy density up to some overall factor.
For static, spherically symmetric solutions, the scalar ansatz is still given by (\ref{phi}) with $w=0$. 

The equations of the model can also be derived from the reduced Lagrangian:
\begin{eqnarray}
\label{Lsph}
L=\frac{1}{16\pi G}L_g-  L_s \ , \qquad {\rm where} \qquad 
L_g=
6\sigma r\left(1-N-\frac{1}{2}r N'\right)=3\sigma m' \ ,
\end{eqnarray}
and
\begin{eqnarray}
&&
L_s=
 \sigma r^3
\bigg [ 
\lambda_0(1-\cos F)
+\frac{\lambda_1}{2} \left(NF'^2+\frac{3}{r^2}\sin^2F \right)
\\
\nonumber
&&
{~~~~~~~~~}
+3\lambda_2\frac{\sin^2F}{r^2} \left(NF'^2+\frac{\sin^2F}{r^2}\right)
+ \lambda_3\frac{\sin^4F}{r^4} \left(3NF'^2+\frac{\sin^2F}{r^2}\right)
\bigg ] \ .
\end{eqnarray}
This form of the system allows us to derive, following
\cite{Heusler:1996ft,Heuslers}
a generalization of the flat spacetime 
virial identity (\ref{virial1}). 
Following the same reasoning as in Section~\ref{secsca}, we find that the
finite energy solutions satisfy the integral relation
\begin{eqnarray}
\nonumber
&&
\int_0^\infty dr~\sigma r^3
\bigg[
\lambda_1\left(NF'^2+\frac{3\sin^2F}{r^2}\right)
+\frac{2m}{r^2}F'^2  \left( \frac{\lambda_1}{2}+3\lambda_2\frac{\sin^2F}{r^2}+3\lambda_3\frac{3\sin^2F}{r^4}	\right. \bigg.
\\
&&{~~~~~~~~~~~~~~~}
\left. \bigg. +4\lambda_0(1-\cos F)
\right)
\bigg]
\label{virial2}
=
2\lambda_3 \int_0^\infty dr \sigma   \frac{\sin^2F}{r}
\left(
          3N F'^2+\frac{ \sin^2F}{r^2}
					\right),
\end{eqnarray}
which clearly shows that nontrivial gravitating solutions with finite mass cannot 
exists in a model without the sextic term.
Indeed, in that case the right hand side would vanish and all terms in the integrand of the left hand side of (\ref{virial2}) would either vanish or be strictly positive, making the equality impossible for a non-trivial configuration. 
Observe that turning on gravity adds an extra attractive term, in addition to those provided by the quadratic and potential terms.

By using the same dimensionless radial coordinate
and rescaling
 as in the 
non-gravitating case 
(together with $m \to m/L^2$),
one finds that the gravitating system possess one extra dimensionless coupling constant
\begin{eqnarray}
\label{alpha}
\alpha^2=4\pi G \lambda_1^{3/2}/\lambda_0^{1/2}~.
\end{eqnarray}

Then the Einstein equations used in the numerics
reduce to (recall that we set
$\lambda_0=\lambda_1=1$ and $\lambda_2=0$, $\lambda_3=1$)
\begin{eqnarray}
\label{eqm}
&&
m'=\frac{4}{3}\alpha^2  r^3
\bigg[
\frac{1}{2} \left(NF'^2+\frac{3\sin^2F}{r^2}\right)
+3\lambda_2 \frac{ \sin^2F}{r^2}\left(NF'^2+\frac{\sin^2F}{r^2}\right)
\\
\nonumber
&&
{~~~~~~~~~~~~~~~~~~~~~~~~~~~~~~~~}
+\lambda_3 \frac{ \sin^4F}{r^4 }\left(3NF'^2+\frac{\sin^2F}{r^2}\right)
+ 1-\cos F
\bigg] \ ,
\\
\label{eqs}
&&
\sigma'=\frac{4}{3}\alpha^2  r \sigma
\left(
\frac{1}{2}+3\lambda_2\frac{\sin^2F}{r^2}+3\lambda_3 \frac{\sin^4 F}{r^4}
\right)F'^2 \ ,
\end{eqnarray}
together with an extra constrain equation.
The function $F(r)$ satisfies the 2nd order equation 
\begin{eqnarray}
\label{eqF}
F''=\frac{\sin F}{N\left(1+6\lambda_2\frac{\sin^2 F}{r^2}+6\lambda_3\frac{\sin^4 F}{r^4}\right)}
\left(
1+\frac{1}{r^2\sin^2F}{\cal P}_1
-\frac{3\lambda_2}{r^4}{\cal P}_2
-\frac{3\lambda_3 \sin^2F}{2r^6}{\cal P}_3
\right)
\end{eqnarray}
 where
\begin{eqnarray}
\nonumber
&&
 {\cal P}_1=\frac{3}{2}\sin(2F)-rF' \left[r N'+N\left(3+\frac{r\sigma'}{\sigma}\right) \right]\ ,
\\
&&
\label{eqFs}
 {\cal P}_2=\cos(3F)+\cos F (2N r^2F'^2-1)+2r  F'\sin F\left[r N'+N\left(1+\frac{r\sigma'}{\sigma}\right) \right] \ ,
\\
&&
\nonumber
 {\cal P}_3=\cos(3F)+\cos F (8N r^2F'^2-1)+4r  F'\sin F\left[r N'+N\left(-1+\frac{r\sigma'}{\sigma}\right) \right] \ .
\end{eqnarray}
%

\vspace*{-0.1cm}
 {\small \hspace*{3.cm}{\it  } }
\begin{figure}[ht]
\hbox to\linewidth{\hss%
	\resizebox{8cm}{6cm}{\includegraphics{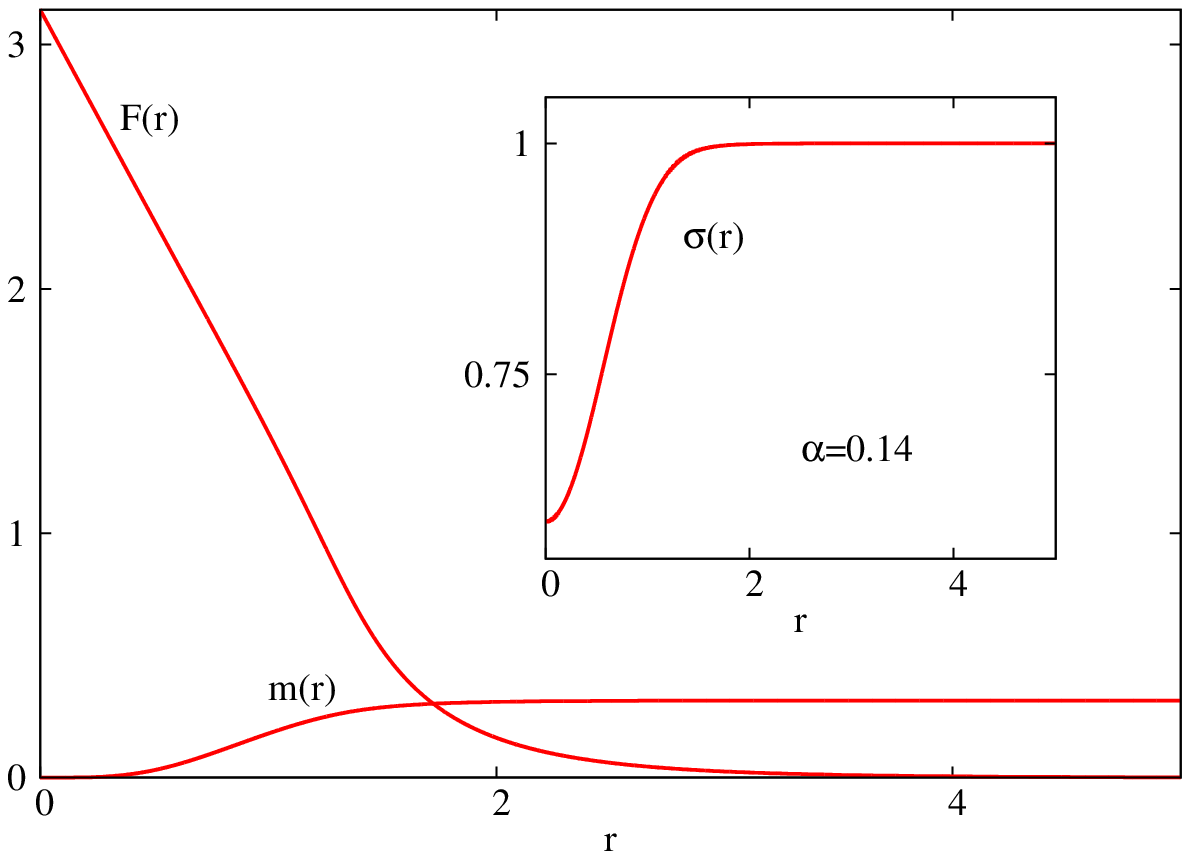}}
\hspace{10mm}%
        \resizebox{8cm}{6cm}{\includegraphics{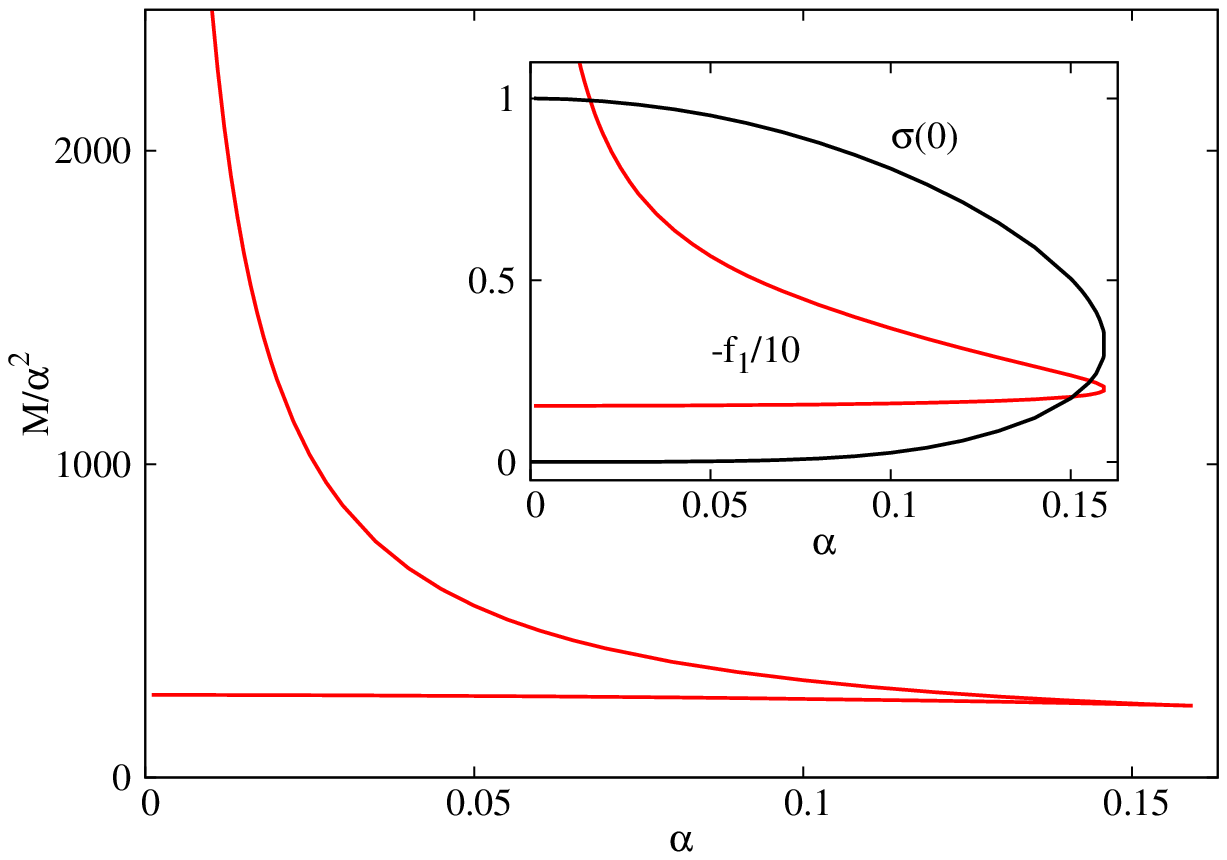}}	
\hss}
 \caption{\small
{\it Left panel:} The  profile functions of a typical 
spherically symmetric Skyrme star. 
{\it Right panel:} The total  mass $M$, the control parameter $f_1$
and the value of the metric function $\sigma$ at the origin 
are shown as a function of coupling constant, 
$\alpha$, for spherically symmetric Skyrme stars.
  }
\end{figure}
 \\
These equations are solved by imposing the boundary conditions\footnote{An approximate form of the solutions compatible with these
conditions can easily be constructed.
For example, the small$-r$ expansion contains two essential parameters, $F(0)$ and $\sigma(0)$. 
}
\begin{eqnarray}
N(0)=1,~~\sigma(0)=\sigma_0>0,~~F(0)=\pi\ ,
\end{eqnarray}
at the origin $(r=0)$,
while at
 infinity, the solutions satisfy
\begin{eqnarray}
N\to 1,~~\sigma\to 1,~~F\to 0 \ .
\end{eqnarray}

The properties of the spherically symmetric 
Skyrme stars can be summarized as follows.
For all studied cases, 
$m(r)$, $\sigma(r)$, and $F(r)$ are monotonic functions of $r$,\footnote{ 
Excited solutions with nodes in the profile of $F$, which would therefore be a non-monotonic function, are likely to exist as well.}
the profile of a typical solution being presented in Figure 3 (left panel).
For small values of $\alpha$ there is a \textit{fundamental branch} of solutions that reduces to the flat space Skyrmion as
$\alpha \to 0$.
When $\alpha$
increases, the mass parameter $M$ decreases, as well as the value $\sigma(0)$.
The  solutions exist up to a maximal
value $\alpha_{max}$ of the parameter $\alpha$.
At the same time, the absolute value of the ``shooting" parameter $f_1$ increases with $\alpha$.
We found that a \textit{secondary branch} of solutions emerges at  $\alpha_{max}$,
extending backwards in $\alpha$ - Figure 3 (right panel).
Along this second branch, both $\sigma(0)$ and
$f_1$ decrease as $\alpha$ decreases, 
while the value of the physical mass $M/\alpha^2$ strongly increases.
Some understanding of the limiting behaviour can be obtained by noticing that
the $\alpha\to 0$ limit can be approached in two different ways, as $G\to 0$
(flat space, first branch)
or as $\lambda_1\to 0$ (second branch).
Then 
we conjecture that
the limiting solution on the upper branch corresponds
to a gravitating model without the ${\cal F}^2$-term in the Skyrme Lagrangian.

\subsection{Spinning stars}

$d=5$ rotating spacetimes 
generically possess two independent angular momenta. Here, however, 
 we focus on configurations with
equal-magnitude angular momenta
which are compatible with the symmetries of the matter energy-momentum tensor we have chosen. 
A suitable metric ansatz reads
(note the existence of a residual gauge freedom
which will be fixed later):
 \begin{eqnarray}
\label{metric-rot}
&&ds^2 = \frac{dr^2}{f(r)}
  + g(r) d\theta^2
+h(r)\sin^2\theta \left[ d \varphi_1 -W(r)dt \right]^2
+h(r)\cos^2\theta \left[ d \varphi_2 -W(r)dt \right]^2 ~~{~~~~~}
\\
\nonumber
&&{~~~~~~}+[g(r)-h(r)]\sin^2\theta \cos^2\theta(d \varphi_1 -d \varphi_2)^2
-b(r) dt^2~.
\end{eqnarray}
For such solutions the isometry group is enhanced from $\mathbb{R}_t\times U(1)^2$ to $\mathbb{R}_t\times U(2)$, where $\mathbb{R}_t$
denotes the time translation. 
This symmetry enhancement allows factorizing the angular
dependence and thus leads to ordinary differential equations.

The angular momentum and energy densities are given by
\begin{eqnarray}
\nonumber
j=\frac{T_{\varphi_1}^t}{\sin^2\theta}=\frac{T_{\varphi_2}^t}{\cos^2\theta}=
\frac{(\om-W)}{b(r)}\sin^2F
\bigg\{
\lambda_1+2\lambda_2 f F'^2
+\frac{2\sin^2F}{g^2}
\left[
\lambda_3 \sin^2F +2g(\lambda_2+\lambda_3 F'^2)
\right]
\bigg\},
\end{eqnarray}
\begin{eqnarray}
\nonumber
\rho=-T_t^t&=&
\lambda_0(1-\cos F)+\frac{1}{2} 
\lambda_1 
\left [
f F'^2+\left(\frac{2}{g}+\frac{1}{h}+\frac{\om^2-W^2}{b}\right)\sin^2F
\right]
\\
\nonumber
&&
+\lambda_2 \sin^2 F 
\left\{
  \left[
  \left(\frac{2}{g}+\frac{1}{h}\right)f
	+\frac{f}{b}(\om^2-W^2) 
	      \right]F'^2
	+
	\left(
	\frac{1}{2g}+\frac{1}{h}+\frac{\om^2-W^2}{b}
	\right)\frac{2\sin^2F}{g}
\right\}
\\
\nonumber
&&
+\lambda_3
\frac{\sin^4 F}{g}
\left[
\left(
\frac{1}{2g}+\frac{1}{h}+\frac{\om^2-W^2}{b}
\right)2f F'^2
+
\left(
          \frac{1}{h}+\frac{\om^2-W^2}{b}
\right)
              \frac{\sin^2F}{g}
\right ]~.
\end{eqnarray}

The complete ansatz, (\ref{metric-rot}) and (\ref{phi}), can be proven to be consistent, and, as a result, the Einstein-Skyrme
 equations reduce
to a set of five ODEs 
(in the numerics, we fix the metric gauge by taking $g(r)  = r^2$).

We seek asymptotically flat solutions,
subject to the following boundary conditions as $r\rightarrow \infty$:
$f=b=g(r)/r^2=1$
and
$F=W=0$.
The total (ADM) mass $M$ and angular momenta $J_1=J_2=J$, are read off from the asymptotic behaviour of the metric functions,
\begin{eqnarray}
\label{asym}
g_{tt} =-1+\frac{8 G M}{3\pi r^2}+\dots,
~~g_{\varphi_1 t}=-\frac{4 GJ}{\pi r^2}\sin^2\theta+\dots,
~~g_{\varphi_2 t}=-\frac{4 G J}{\pi r^2}\cos^2\theta+\dots~.
\end{eqnarray}
The behaviour of the metric functions 
at the origin
is
$f=1$, 
$b=b_0>0$,
$g(0)=0$,
together with
$W=W_0>0$,
while $F=\pi$,
as in the probe limit.

In Figure 4, we display the profiles of the a typical spinning Skyrme star.
The corresponding distribution for energy and angular momentum densities
look similar to those shown in Figure 1 for the probe limit.
The  mass/angular momentum of solutions $vs.$ frequency  are given as a limiting curve in Figure 11,
for a particular value of $\alpha$. 
Note that similar to the probe limit, both quantities monotonically increase with $\om$.
Also, in the limit $\om \to 1$, the solutions 
are still localized, without any special features, while they disappear for $\om > 1$. 

Finally we emphasise that all solitons in this work, gravitating or otherwise,
possess $unit$ topological charge, as expected.

 {\small \hspace*{3.cm}{\it  } }
\begin{figure}[t!]
\hbox to\linewidth{\hss%
	\resizebox{8cm}{6cm}{\includegraphics{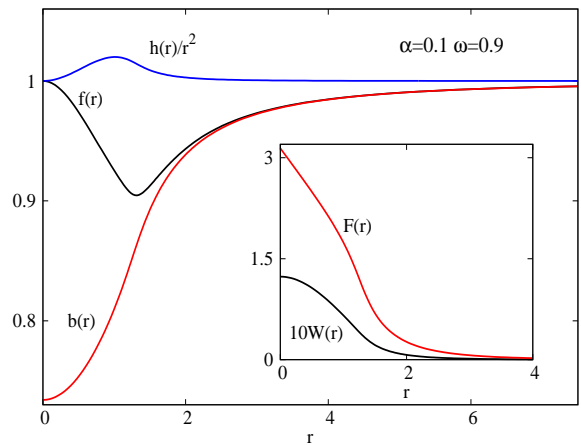}}
\hss}
\caption{\small 
The profile functions of a typical spinning Skyrme star.
}
\end{figure}

\section{BHs with Skyrme hair}

BH generalizations are generically found for any regular
solitonic-like gravitating configuration, at least for small values of the horizon radius $r_H$.
In this Section we shall show this trend remains true for the Einstein-Skyrme model discussed herein,  as  confirmed by our numerical results.

\subsection{Spherically BHs}

\subsubsection{The probe limit -- Skyrmions on a Schwarzschild BH background}

Similarly to the $d=4$ case, it is useful to consider first the
probe limit and solve the Skyrme equations on a $d=5$  Schwarzschild-Tangherlini BH background~\cite{Tangherlini:1963bw}.
The corresponding line element is given by (\ref{sph-metric}) with 
$N(r)=1-r_H^2/r^2$ and $\sigma(r)=1$, where $r_H>0$ the event horizon radius. 
The approximate form of the solution close to the horizon reads
\begin{eqnarray}
F(r)=f_0+f_1(r-r_H)+\mathcal{O}(r-r_H)^2,
\end{eqnarray}
in terms of the ``shooting" parameter $f_0$, with
\begin{eqnarray}
f_1= \frac{\sin(f_0)[\lambda_0 r_H^6 +3\cos(f_0)(r_H^4\lambda_1 +2\lambda_3\sin^4 f_0)]}
{2r_H(r_H^4\lambda_1 +6 \lambda_3 \sin^2 f_0)}~.
\end{eqnarray}
 The mass of the solutions is
still computed from (\ref{MJ}),
with the corresponding curved spacetime expressions
and $r=r_H$ as a lower bound in the integral
(the same holds for a MP background).

 {\small \hspace*{3.cm}{\it  } }
\begin{figure}[t!]
\hbox to\linewidth{\hss%
	\resizebox{8cm}{6cm}{\includegraphics{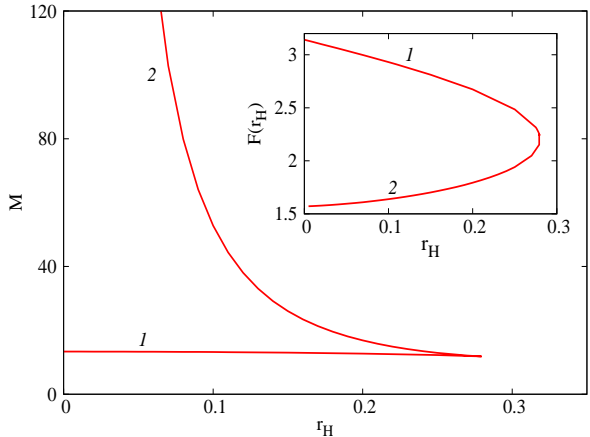}}
\hss}
\caption{\small 
 The  mass $M$  and the control parameter $F(r_H)$ are shown
 as functions of the horizon radius $r_H$,
 for Skyrme probe solutions on a Schwarzschild BH background.}
\end{figure}

The  results of the numerical integration are shown in Figure 5 
(note that the typical profile of the function $F(r)$
is similar to that exhibited in the gravitating case, $cf.$ Figure 6).
One can see that the solutions exist up to a maximal horizon radius $r_H$
of the Schwarzschild background, with a double branch structure
for a range of $r_H$.
The solutions of the fundamental branch (with label 1 in  Figure 5) terminate in the flat spacetime solitons as $r_H\to 0$,
with $F(r_H)\to \pi$ in that limit.
Along this branch, the mass of the solutions decreases with increasing $r_H$.
The second branch   (with label 2 in  Figure 5) starts at $r_H^{(max)}$
and continues again to $r_H\to 0$,
in which limit however, the mass of the solution $M$
diverges, while $F(r_H)\to \pi/2$.

We remark that the Skyrme solutions on a spacetime geometry with an event horizon
possess a non-integer topological charge,
\begin{eqnarray}
\label{B-BH}
B=\left(1+2\cos^2\left[\frac{1}{2}F(r_H)\right] \right) \sin^4\left[\frac{1}{2}F(r_H)\right] \ , 
\end{eqnarray}
belonging to the interval $1/2 < B\leq 1$.
In fact, the above expression holds for all BH solutions in this work, 
including the rotating ones.

\begin{figure}[ht]
\hbox to\linewidth{\hss%
	\resizebox{8cm}{6cm}{\includegraphics{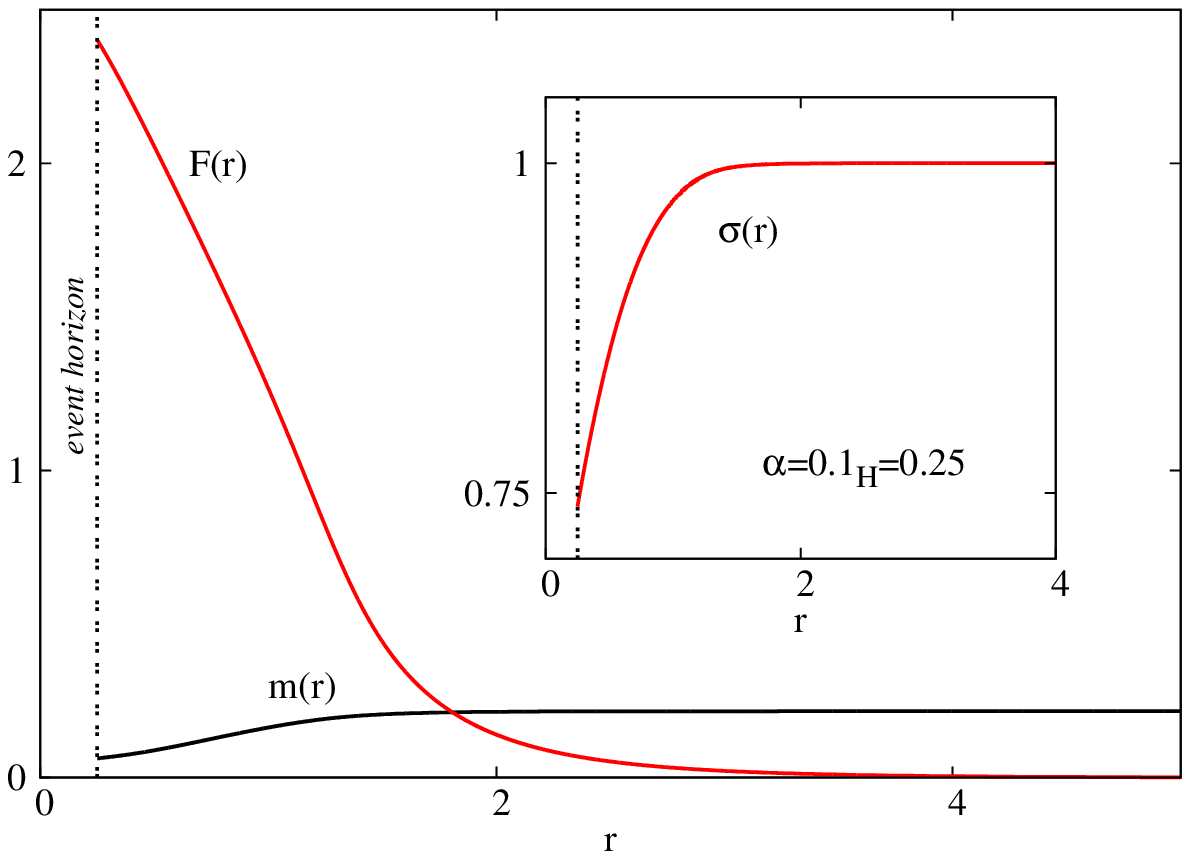}}
\hspace{10mm}%
        \resizebox{8cm}{6cm}{\includegraphics{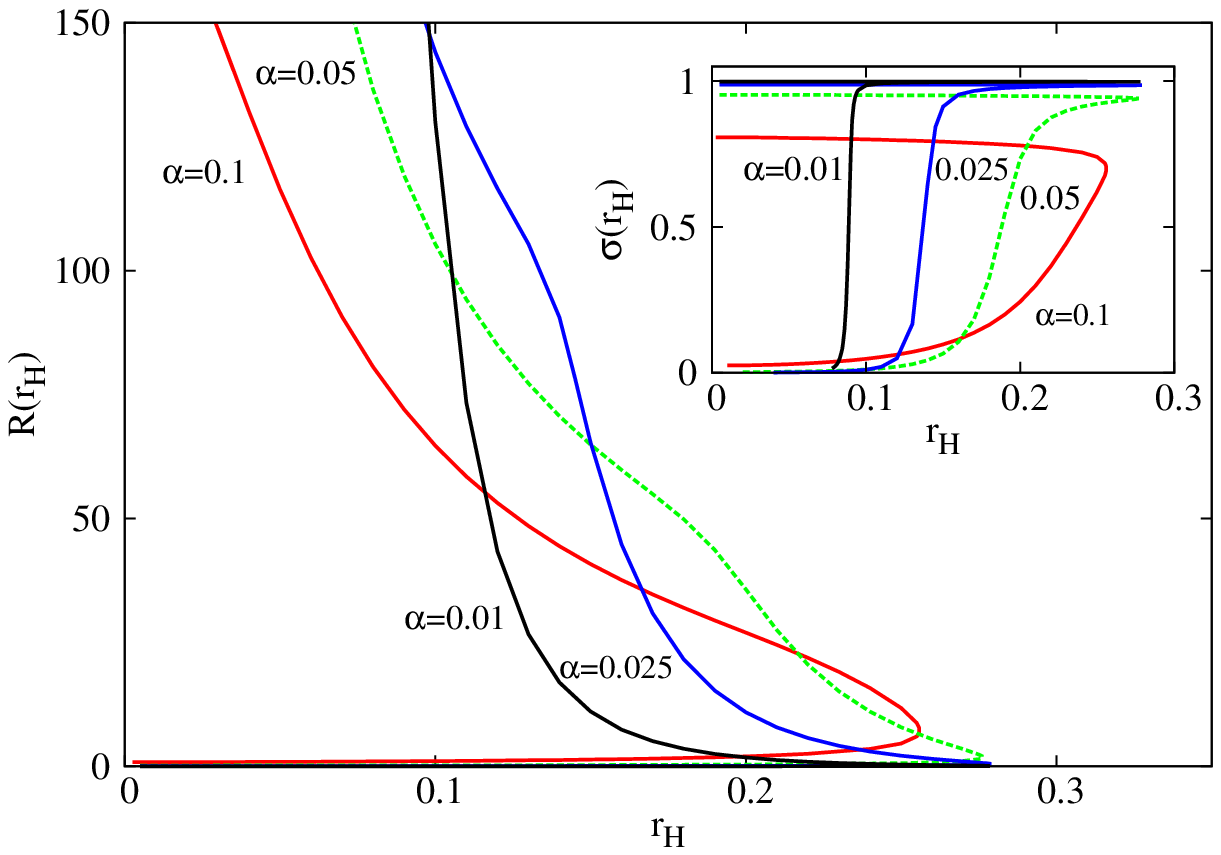}}	
\hss}
\caption{\small 
{\it Left panel:} 
The  profile functions of a typical
spherically symmetric BH with Skyrme hair. 
{\it Right panel:} 
 The  values, at the horizon, of the Ricci scalar $R$ and of the metric function $\sigma$ are shown as functions of the horizon radius $r_H$, 
 for several values of $ \alpha$.
}
 \end{figure}

\begin{figure}[ht]
\hbox to\linewidth{\hss%
	\resizebox{8cm}{6cm}{\includegraphics{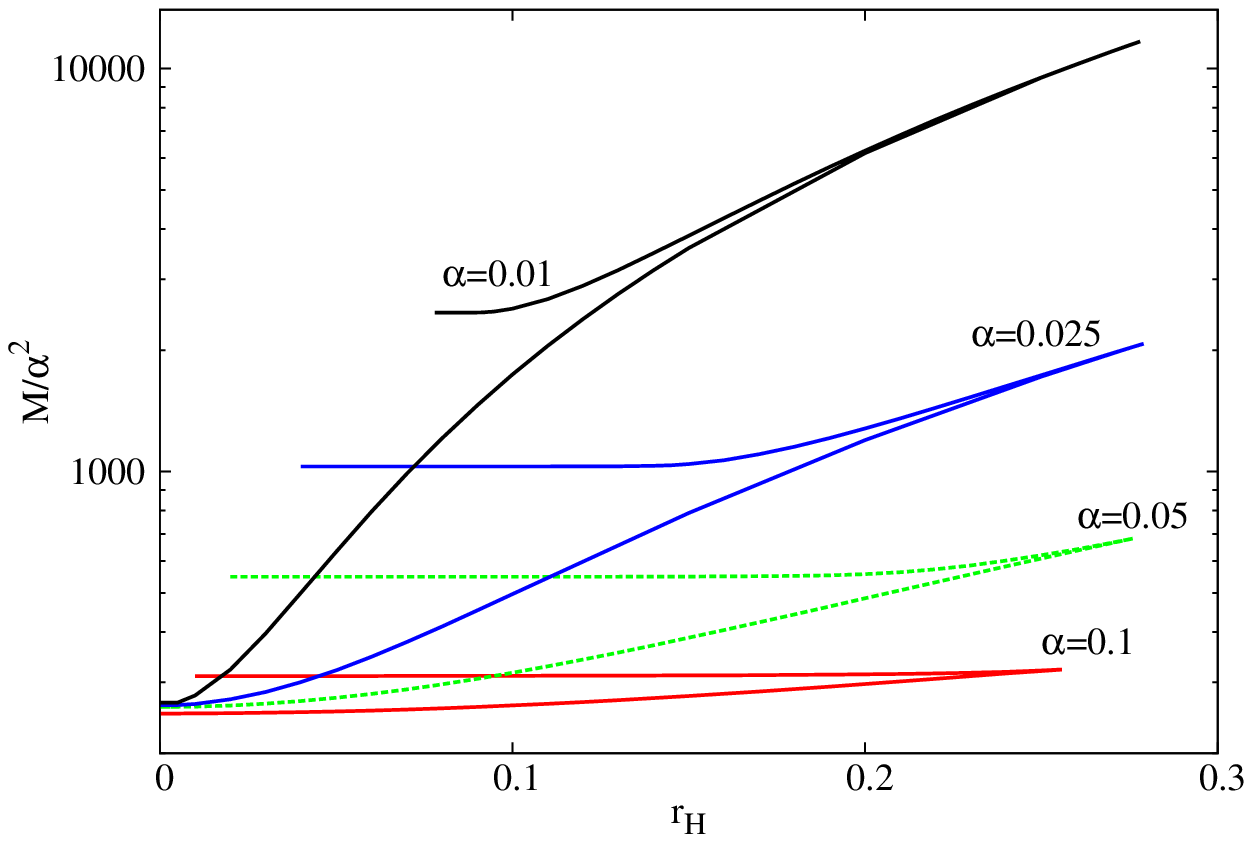}}
\hspace{10mm}%
        \resizebox{8cm}{6cm}{\includegraphics{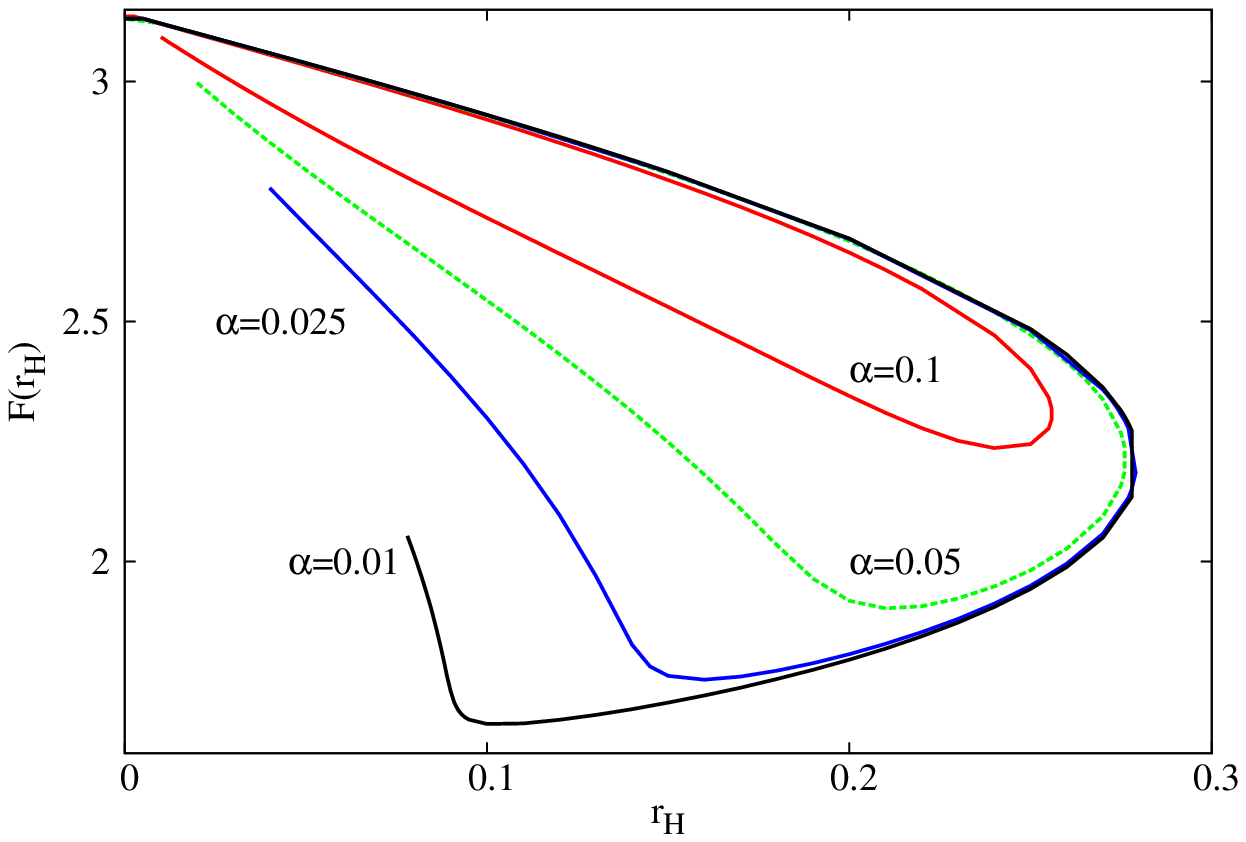}}	
\hss}
\caption{\small 
 The  mass $M$ (left panel) and the control parameter $F(r_H)$ (right panel) are shown as functions of the horizon radius $r_H$, 
 for several values of $ \alpha$.
}
 \end{figure}

\subsubsection{Including backreaction}

The inclusion of gravity effects is straightforward.
The BH solutions are constructed 
within the same ansatz used for solitons. 
They satisfy the following set of boundary conditions at the horizon (which is located at $r=r_H$,
 with $0<r_H\leqslant r<\infty$)
\begin{eqnarray}
N(r_H)=0\ ,~~\sigma(r_H)=\sigma_H>0\ ,~~F(r_H)=f_0~.
\end{eqnarray}
The far field behaviour is similar to that in the solitonic case.
We note that these BHs possess a Hawking temperature $T_H$ and a horizon area $A_H$, which read
\begin{eqnarray}
 T_H= \frac{1}{4\pi}N'(r_H)\sigma(r_H) \ ,~~A_H=2\pi^2 r_H^3\ .
\end{eqnarray}

The profile of a typical BH solution is shown in Figure 6.
The behaviour of the solutions as a function of $r_H$ is presented in Figure 7, 
for several  values of the coupling constant $\alpha$.
The properties of the spherically symmetric BHs can be summarized as follows.
Starting from any regular solution, $i.e.$ a Skyrme star, with a given $\alpha$ and increasing the event horizon radius,
one finds a first branch of solutions which extends to a maximal value
$r_{H}^{(max)}$.
This maximal value decreases with increasing $\alpha$.
This branch is the backreacting counterpart of the corresponding one
in the probe limit.
The Hawking temperature and the value of $F(r_H)$ decrease along this branch, while the mass
parameter increases; 
however, the variations of the mass and of $\sigma(r_H)$ are relatively small.

Extending backwards in $r_H$,
we find a second branch of solutions.
This second branch extends up to a critical value of horizon radius $r_H^{(cr)}$
where an essential singularity seems to occur. 
An understanding of the limiting solutions requires a reformulation of the problem with a different coordinate system
\cite{Breitenlohner:2005hx}
which is beyond the scope of this work.
Here we note that the value of $\sigma(r_H)$ on this branch decreases drastically 
and appears to vanish as $r_H\to r_H^{(cr)}$.
As a result, the Ricci scalar evaluated at the horizon strongly increases in that limit.
However, the  mass remains finite, while
the Hawking temperature goes to zero.
The profile of the $F$-function does not exhibit any special features in the limit, starting always
at some value $F(r_H)>\pi/2$.
This special behaviour on the second branch 
can partially be understood as a manifestation of the divergent behaviour
we have noticed in the probe limit. 

We also mention that
for the region of the parameter space where 
two different solutions exist with the same mass,
the event horizon area ($i.e.$ the entropy) is always
maximized by the fundamental branch of the BH, see Figure 7 (left panel).
Thus we expect the upper branch solutions to be always unstable.

\subsection{Spinning BHs}

\subsubsection{The probe limit - Skyrmions on a Myers-Perry BH background}

The static hairy BHs we have just described possess 
rotating generalizations.
However, before considering solutions of the full Einstein-Skyrme
system, it is again useful to consider first the probe limit
and to solve the matter field equations on a spinning BH background.
The corresponding BH is a $d=5$ Myers-Perry (MP) solution~\cite{Myers:1986un}
with two equal-magnitude angular momenta.
Such a BH can be expressed as a particular case of the  ansatz (\ref{metric-rot}), by taking
\begin{eqnarray}
\label{MP}
\nonumber
 &&f(r)=1-\frac{\left(\frac{r_H}{r}\right)^2}{1-r_H^2\Omega^2_H}
 +\frac{r_H^2\Omega_H^2}{1-r_H^2\Omega_H^2}\left(\frac{r_H}{r}\right)^4\ ,
\qquad 
 h(r)=r^2
 \left[
 1+\left(\frac{r_H}{r}\right)^4\frac{r_H^2\Omega_H^2}{1-r_H^2\Omega_H^2}
  \right]\ ,
\\
\nonumber
&&b(r)=1
 -
 \frac{\left(\frac{r_H}{r}\right)^2}{1-\left[1-(\frac{r_H}{r})^4\right]r_H^2\Omega_H^2}\ , \qquad g(r)=r^2\ ,
 \qquad W(r)=\frac{ \Omega_H \left(\frac{r_H}{r}\right)^4}{1-\left[1-\left(\frac{r_H}{r}\right)^4\right]r_H^2\Omega_H^2} \ ,
\end{eqnarray}
and it is parameterised 
in terms of the event horizon radius $r_H$
and the horizon angular velocity $\Omega_H$,
which are the control parameters in our numerical approach. 
For completeness, we include the expression of quantities 
which enter the thermodynamics of a MP BH (with $G=1$):
\begin{eqnarray}
&&
\label{MP-quant1}
M^{(MP)}=\frac{3\pi r_H^2}{8(1-\Omega_H^2 r_H^2)}\ , \qquad
J_1^{(MP)}=J_2^{(MP)}=J^{(MP)}=\frac{\pi \Omega_H r_H^4}{4(1-\Omega_H^2 r_H^2)}\ , 
\\
&&
\nonumber
T_H^{(MP)}=\frac{1}{2\pi r_H}\frac{1-2\Omega_H^2 r_H^2}{\sqrt{1-\Omega_H^2 r_H^2}}\ , \qquad
A_H^{(MP)}=\frac{2\pi^2 r_H^3}{\sqrt{1-\Omega_H^2 r_H^2}} \ .
\end{eqnarray}
Here it is important to note the
existence of a maximal size of the horizon radius $r_H$
for a given  value of  horizon angular velocity $\Omega_H$.
This corresponds to a  zero temperature BH with
$r_H^{(max)}=1/(\sqrt{2}\Omega_H) $.
There, the reduced angular momentum  
(\ref{red-j}) approaches its maximal value
$\bar j_{MP}=1$.

At infinity, the decay of the field is still given by (\ref{infS}), such that
$
F\to 0.
$
Remarkably, the assumption of existence of a power series expansion 
of $F(r)$
as $r\to r_H$ 
implies that,
similar to other hairy rotating BH solutions~\cite{Herdeiro:2014goa,Brihaye:2014nba,Herdeiro:2015kha},
 the {\it synchronization condition}
\begin{eqnarray}
\label{synch}
\om=\Omega_H
\end{eqnarray}
necessarily holds. This condition is also implied by the regularity of 
energy and angular momentum densities as $r\to r_H$.
Then the function $F(r)$ possess an approximate solution
near the horizon (with $\lambda_2=0$), which, up to order $\mathcal{O}(r-r_H)$ reads:
\begin{equation}
\label{F-MP}
\small
F(r)=f_0+\frac{(1-r_H^2 \om^2)
\left[-2\lambda_0 \sin f_0
+\lambda_1 r_H^4 (r_H^2 \om^2-1)\sin(2f_0)
+12\lambda_3(r_H^2 \om^2-1)\cos f_0 \sin f_0^5 \right]}
{
4r_H(2r_H^2 \om^2-1)[r_H^2\lambda_1+2\lambda_3(3-2r_H^2\om^2)\sin^4 f_0] 
}
(r-r_H) \ ,
\end{equation}
all higher order coefficients being determined by $F(r_H)>0$.

The profiles of a typical solution on a given MP background are shown in Figure 8. 
The dependence of the properties of the solutions on the horizon size, as given by $A_H^{(MP)}$
and reduced angular momentum $j_{(MP)}$, is shown in Figure 9.
The basic picture found in the static case is still valid here,
with the existence of two branches of solutions for a given BH background.
The fundamental branch emerges from the flat spacetime solitons,
while the  mass and angular momentum of the solutions
appear to diverge as the flat spacetime limit ($r_H\to 0$)
is approached, along the second branch.

 {\small \hspace*{3.cm}{\it  } }
\begin{figure}[t!]
\hbox to\linewidth{\hss%
	\resizebox{8cm}{6cm}{\includegraphics{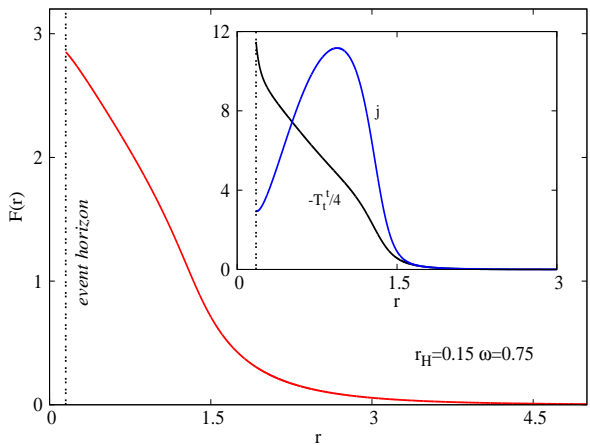}}
\hss}
\caption{\small 
The  profile of a  typical Skyrme test field solution on a given MP background is shown together with 
the corresponding energy and angular momentum densities.
}
\end{figure}

\begin{figure}[ht]
\hbox to\linewidth{\hss%
	\resizebox{8cm}{6cm}{\includegraphics{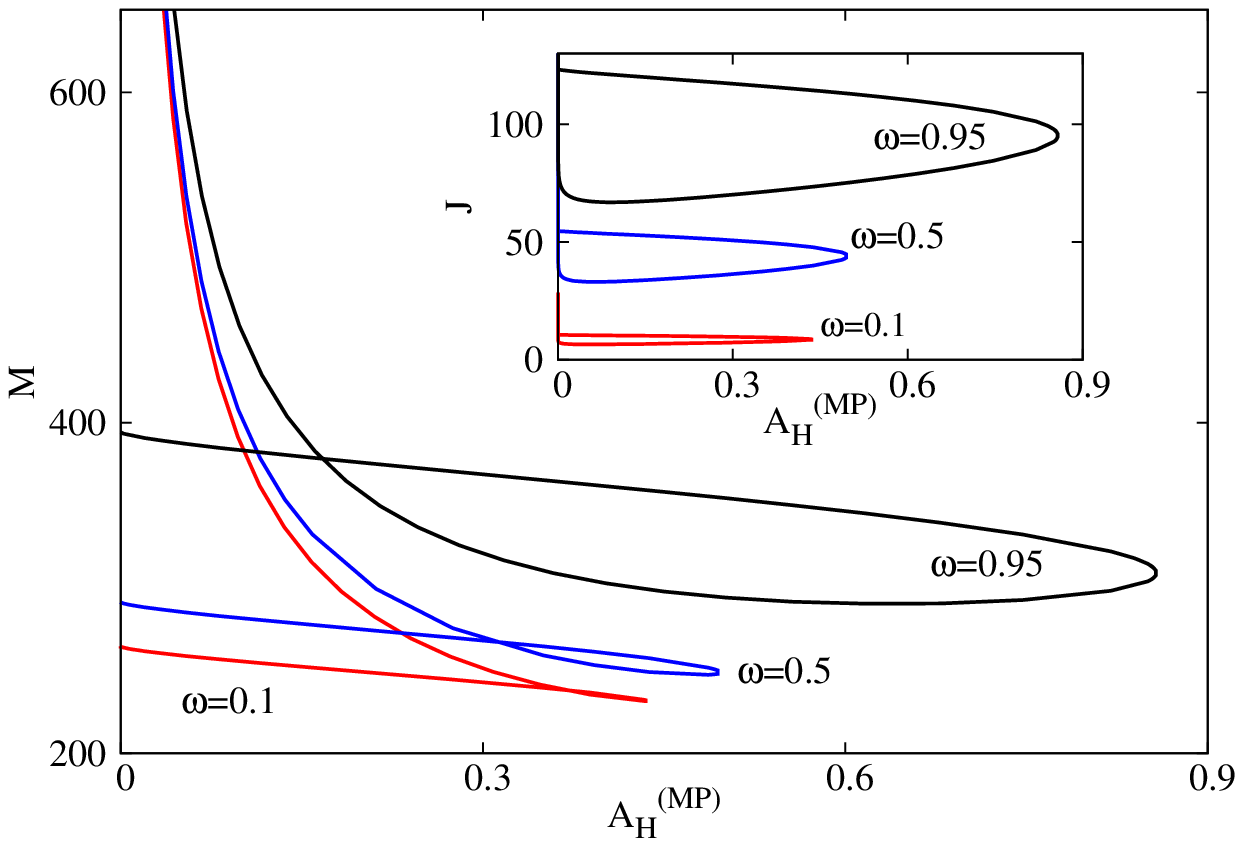}}
\hspace{10mm}%
        \resizebox{8cm}{6cm}{\includegraphics{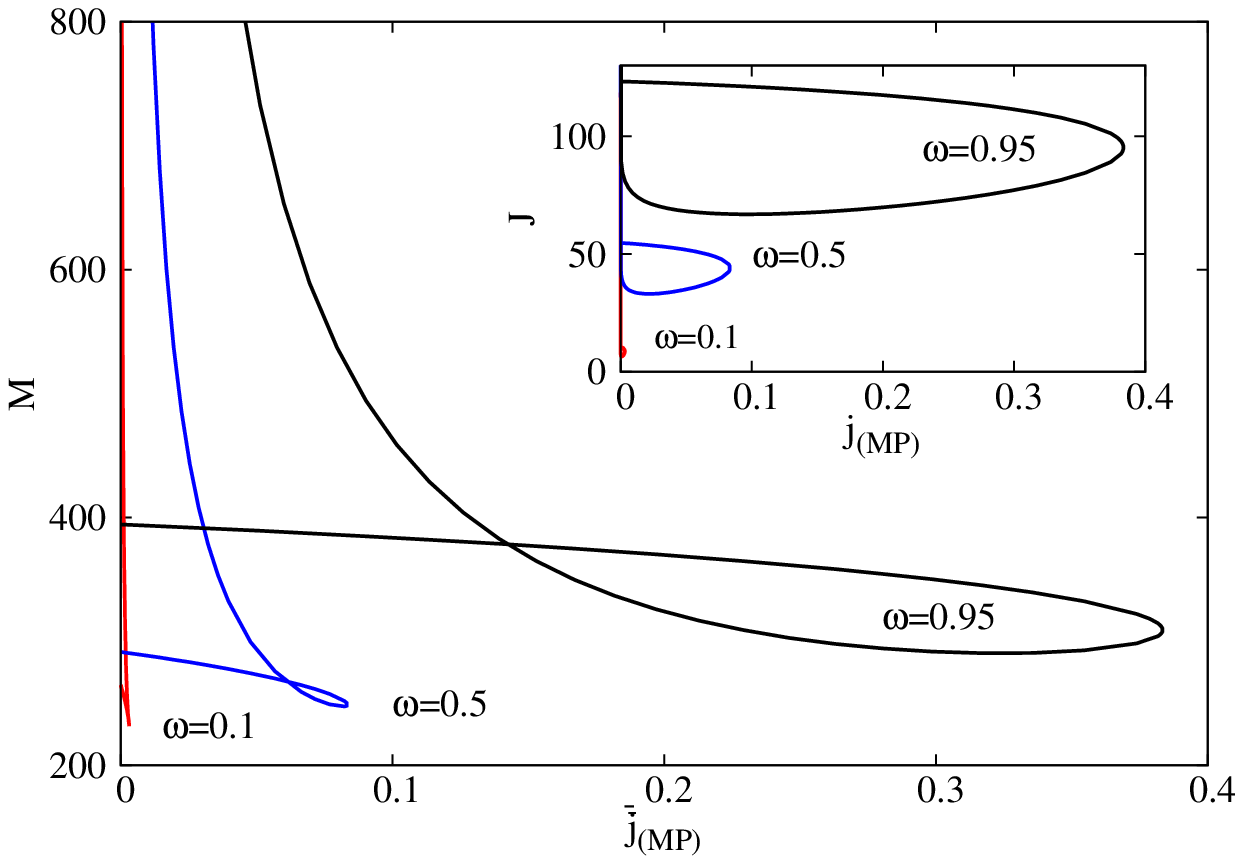}}	
\hss}
\caption{\small 
The  mass and angular momentum of  Skyrme test field solutions on a MP background are shown as a function of horizon area
and of the reduced MP angular momentum.
}
 \end{figure}

We remark that no Skyrme solutions exist on a fast rotating MP background, $i.e.$ with 
$j_{(MP)}$  close to unity. Also, one can see that the mass branches of spinning solutions exhibit a 'loop', 
when considered as a function of horizon properties (or reduced angular momentum) 
of the MP background.

\subsubsection{Including backreaction}
 
The configurations described in the last subsection survive when taking into account their  
backreaction on the spacetime geometry.
The spinning BHs
 are constructed within the same ansatz as for the spinning
solitons discussed in Section 3.2.2 (in particular we set again $g(r)=r^2$).
However, they possess an horizon which is a squashed $S^3$ sphere. 
The horizon resides at the constant value of
the radial coordinate $r=r_H>0$, and it is characterized by $f(r_H)=b(r_H) =  0$. 
Restricting to nonextremal solutions, the following expansion of the metric functions
holds near the event horizon: 
\begin{eqnarray}
\label{c1}
f(r)=f_1(r-r_H)+f_2(r-r_H)^2+\mathcal{O}(r-r_H)^3,~~h(r)=h_H+h_2(r-r_H)+  \mathcal{O}(r-r_H)^2,~~{~~}
\\
\nonumber
b(r)=b_1(r-r_H)+b_2(r-r_H)^2+ \mathcal{O}(r-r_H)^3,~~W(r)=\Omega_H+\om_1(r-r_H)+ \mathcal{O}(r-r_H)^2,~{~ }
\end{eqnarray}  
while $F(r)=f_0+f_1(r-r_H)+\dots$.
For a given event horizon radius, 
the essential parameters characterizing the event horizon
are $f_1,~b_1$,~$h_H$,~$\Omega_H$ and $\om_1$ (with $f_1>0,~b_1>0$),
which fix all higher order coefficients in (\ref{c1}).
The construction of the approximate near-horizon solution shows that,
as expected, the synchronization
condition
(\ref{synch})
still holds in the backreacting case.

\begin{figure}[ht]
\hbox to\linewidth{\hss%
	\resizebox{8cm}{6cm}{\includegraphics{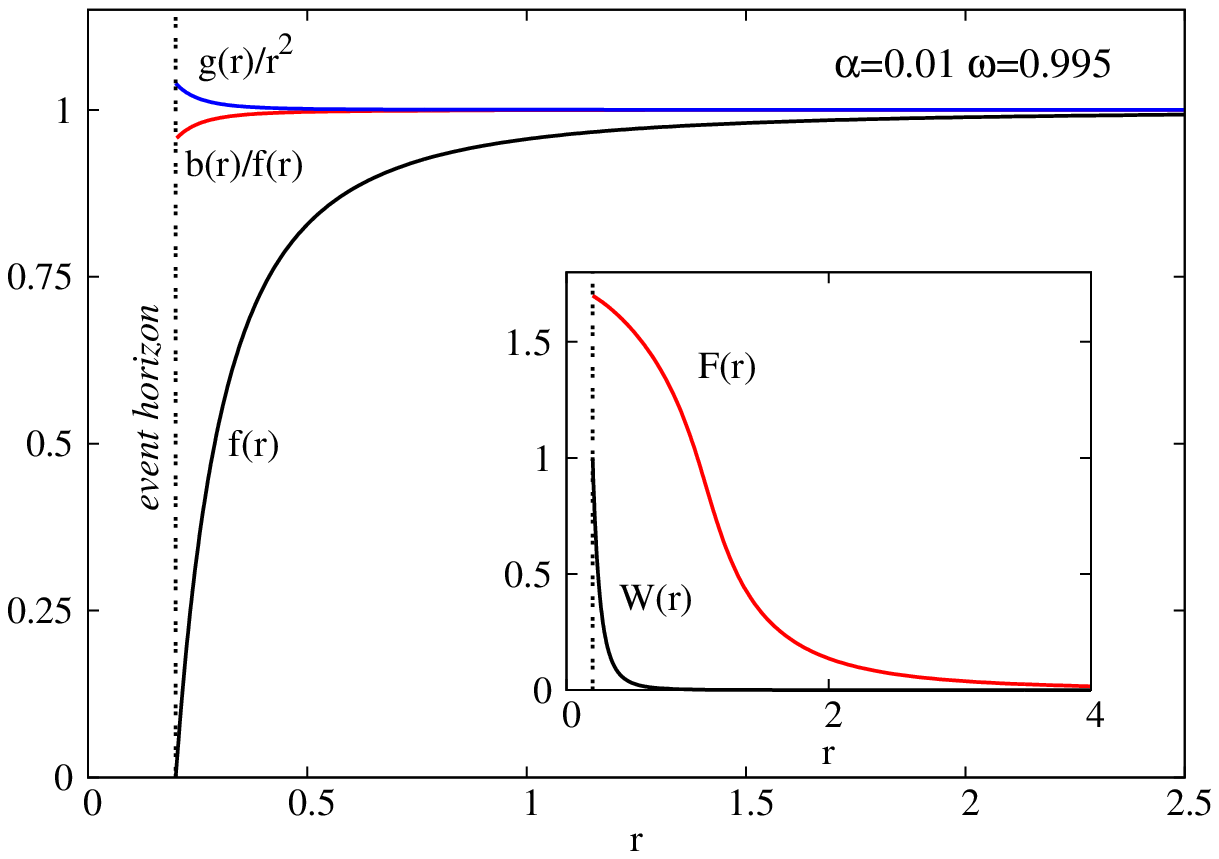}}
\hspace{10mm}%
        \resizebox{8cm}{6cm}{\includegraphics{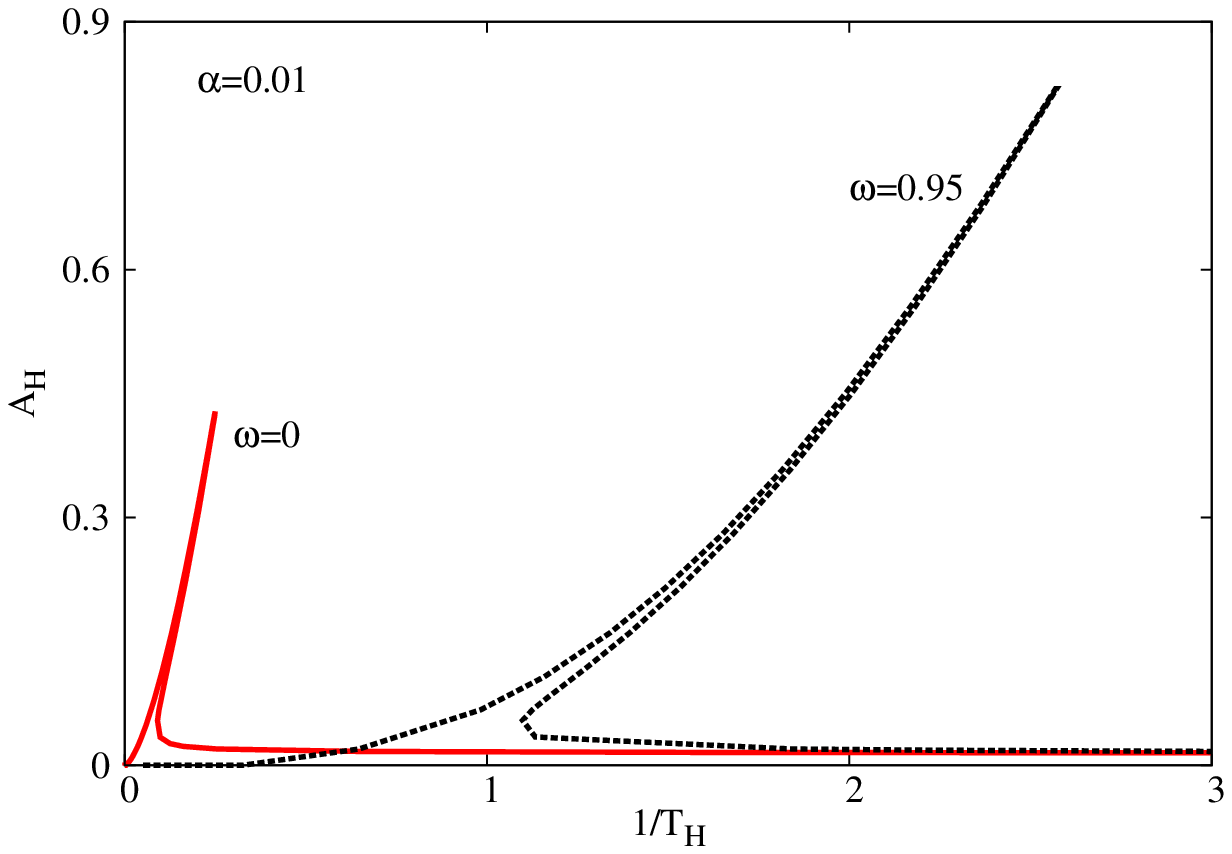}}	
\hss}
\caption{\small 
{\it Left panel:} 
The profile functions of a typical spinning BH with Skyrme hair.
{\it Right panel:} The (horizon area-temperature) diagram is shown 
for two sets of solutions.
}
 \end{figure}

As for a general MP BH,
the (constant) horizon angular velocity $\Omega_H$
is defined in terms of the Killing vector
$\chi=\partial/\partial_t+
 \Omega_1 \partial/\partial \varphi_1 + \Omega_2 \partial/\partial \varphi_2 $
which is null at the horizon.
For the solutions within the ansatz (\ref{metric-rot}), the 
horizon angular velocities are equal, $\Omega_1=\Omega_2=\Omega_H$.
The Hawking temperature  $T_H$  
and  the  area $A_H$ of these BHs are fixed by the near horizon data in (\ref{c1}), with
\begin{eqnarray} 
\label{Temp-rot} 
  T_H=\frac{\sqrt{b_1f_1}}{4\pi},~~A_H= 2\pi^2 \sqrt{h_H}  r_H^2~,
\end{eqnarray} 
while their mass and angular momentum 
are read from the far field expansion (\ref{asym}). 
As usual, 
the temperature, horizon area and the global charges $M,J$ are related through the Smarr mass formula
\begin{eqnarray} 
\label{Smarr} 
\frac{2}{3}M=\frac{1}{4}T_HA_H+2\Omega_H(J-J_{(S)})+\frac{2}{3}M_{(S)}~,
\end{eqnarray}
where $M_{(S)}$ and $J_{(S)}$
are the mass and angular momentum stored in the Skyrme field
outside the horizon,
\begin{eqnarray} 
\label{Smarr2} 
M_{(S)}=-\frac{3}{2}\int_\sigma \sqrt{-g} dr d\theta d\varphi_1 d\varphi_1 
\left(
T_t^t-\frac{1}{3} T_i^i
\right),~~
J_{(S)}= \int_\sigma \sqrt{-g} dr d\theta d\varphi_1 d\varphi_1 
T_{\varphi_i}^t~.
\end{eqnarray}

In our approach, the input parameters 
are  the coupling constant $\alpha$,
the event horizon radius $r_H$ and the horizon angular velocity
$\Omega_H$ (or equivalently, the field frequency $\om$).
Physical quantities characterizing the solutions are then extracted 
from the numerical solutions.  

The profiles of a typical spinning BH with Skyrme hair
are shown in Figure 10 (left panel).
Some basic properties of the solutions are similar to those found in the probe limit.
 For example, the
scalar field is always spatially localized within the vicinity of the horizon,
the distribution of mass and angular momenta densities being similar to that in Figure 8.

The emerging global picture can be summarized as follows.
  For all values of
the parameters
$\alpha,r_H$
 that we have considered,  the static BHs are
continuously deformed while increasing gradually the parameter $\om$.
Similarly to the solitonic case, the solutions stop to
exist for $\om>1$.
Moreover, all BHs studied so far have a reduced angular momentum
(\ref{red-j})
much smaller than one. 
 
\begin{figure}[ht]
\hbox to\linewidth{\hss%
	\resizebox{8cm}{6cm}{\includegraphics{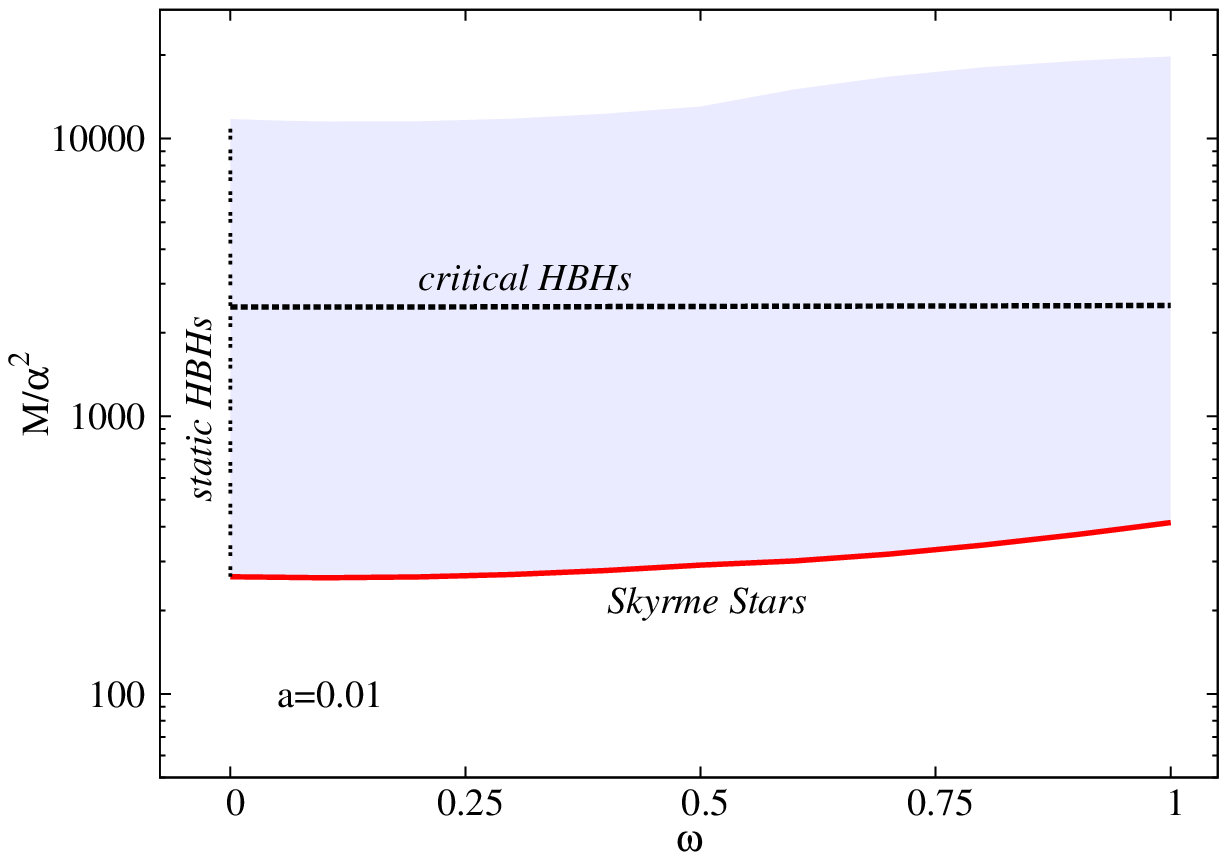}}
\hspace{10mm}%
        \resizebox{8cm}{6cm}{\includegraphics{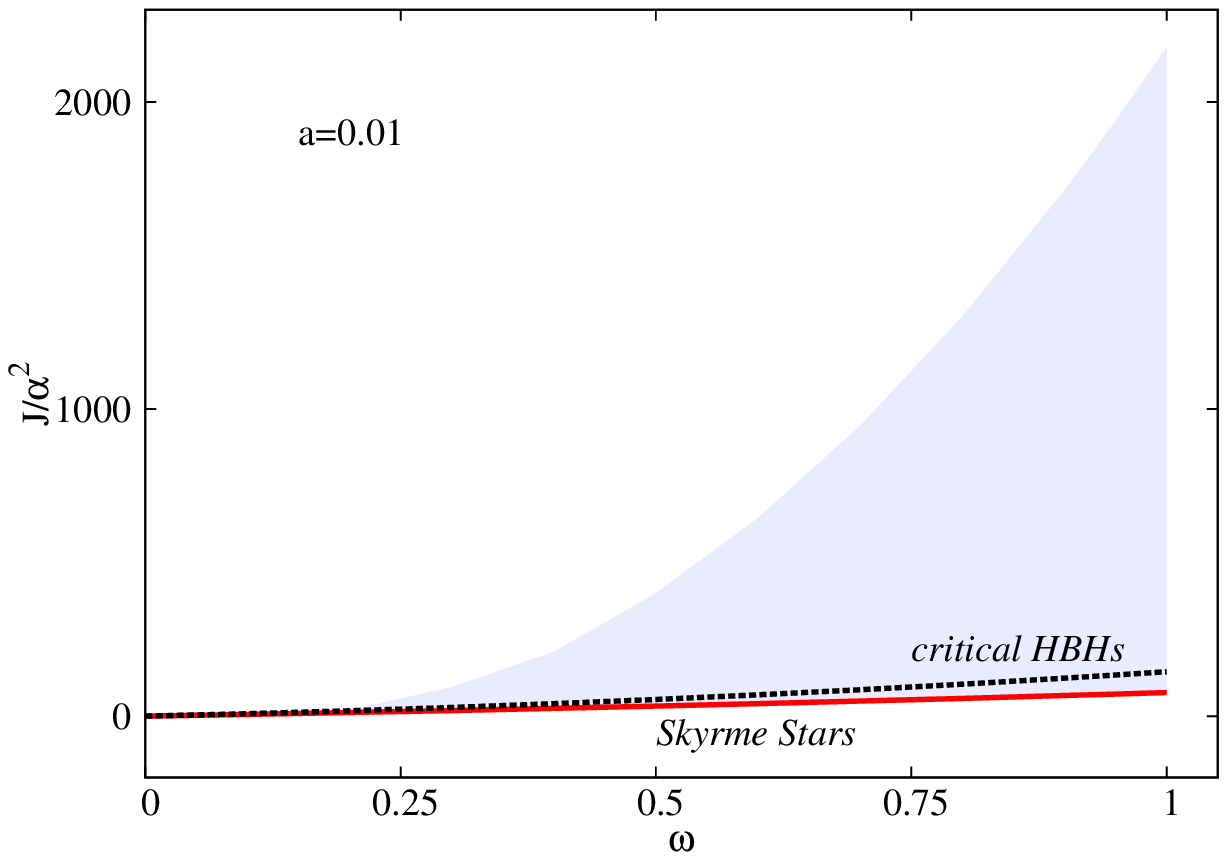}}	
\hss}
\caption{\small 
   The domain of existence of the spinning BHs
	with Skyrme hair is shown in a mass-frequency diagram (left panel) and 
	as an angular momentum-frequency diagram (right panel).
}
 \end{figure}

When taking instead a fixed value of $\om$ and varying the horizon parameter  $r_H$,
our results show that, for any $0\leqslant \om\leqslant 1$, a double branch structure of solutions exists, 
characterized by two
particular values of the horizon radius $r_H$. 
The first (or main) branch exists for $0\leqslant  r_H \leqslant r_H^{(max)}$,
emerging from the corresponding (gravitating)
Skyrme soliton  in the limit $r_H\to 0$.
The second branch exist for $r_H^{(c)}\leqslant  r_H \leqslant r_H^{(max)}$
approaching a critical configuration as $r_H\to r_H^{(c)}$.
This critical solution possesses finite global charges,
 a nonzero horizon area while its Hawking temperature vanishes.
However, it inherits the pathologies of the static limit,
$e.g.$
the  Ricci scalar appears to be unbounded on the horizon.
 The dependence of the horizon area $A_H$ nf the (inverse of the) temperature $T_H$ is
shown in Figure 10 (right panel),
where we compare the results for static solutions with those 
for BHs close to the maximal value of $\om$.
One notices that the horizon size remains finite as the critical solution is approached. 
 
In Figure 11 we exhibit the domain of existence of hairy BHs, in a $M(\om)$ (and $J(\om)$)  diagram
for $\alpha=0.01$, the only value of the coupling constant we have investigated in a systematic way.
This domain, in the $M$-$\omega$ diagram, has an almost rectangular shape, and is delimited by four curves: the set of static BHs ($\om=0$),
the set of Skyrme stars, 
the set of maximal mass solutions, and finally the  set of maximally rotating  BHs with $\om=1$.

\section{Conclusions}

In this work we have considered an extension of the Skyrme model to 
five spacetime dimensions and investigated the basic properties of its codimension-1
solutions.
 
Concerning the model, two salient properties of gravitating Skyrme systems in $d=3+1$, 
can be used to motivate its study. 
The first is that the solutions feature BHs hair and the 
second is that in the gravity decoupling limit the solutions are $topologically \ stable$. In $3+1$
dimensions, gravitating Skyrmions share the the first property
with gravitating Yang-Mills~\cite{Bartnik:1988am}, which support hair but in the flat spacetime limit disappear. 
The second property they share with that gravitating Yang-Mills--Higgs~\cite{Lee:1991vy,Breitenlohner:1991aa,Breitenlohner:1994di}
system which supports topologically stable monopoles. The latter persisit in the gravity decoupling limit, whence one notes the closer similarity of gravitating Skyrmions with monopoles in $3+1$ dimensions.

In spacetime dimensions higher than $d=3+1$, the situation is rather more complicated because of the restriction set by the Derrick scaling~\cite{Derrick:1964ww} requirement for the finiteness of the energy, as can be seen from the
nonexistence of finite energy gravitating solutions of the usual quadratic Yang-Mills system in $4+1$ dimensions~\cite{Volkov:2001tb}. In the Yang-Mills (YM) case, higher order
YM densities, $e.g.$ extended YM terms (eYM) like $F(2p)^2$ 
must be included to satisfy this requirement\footnote{As found in 
\cite{Radu:2011ip},
\cite{Herdeiro:2017oxy},
the inclusion of a $F(4)^2$ term in the Yang-Mills action 
(which is optional in this case)
leads to a variety of new features also for $d=4$,
$e.g.$ the existence of stable hairy non-Abelian black holes.}. 
In $d=6$ and $7$, this was done in \cite{Brihaye:2002hr} by adding\footnote{In these dimensions, 
it is possible to add the $F(6)^2$ eYM term too
but that was eschewed. More interestingly in $d=6$, there exist topologically stable eYM instantons 
so that those solutions persists in the gravity decoupling limit. This feature persists in all even $d=2n\geqslant 4$.}
the $F(4)^2$ eYM term to the usual $F(2)^2$. Subsequently the solutions
for the same model as in \cite{Brihaye:2002hr} were constructed in \cite{Brihaye:2002jg} for $d=5$, which displayed some peculiar features that we have encountered in the $d=5$ Skyrme case at hand. Not surprisingly the same holds for
gravitating Skyrme systems in dimensions greater than $4+1$, namely including higher order kinetic (Skyrme) terms, as we have done in this paper. But unlike in the YM case, where the gravity decoupling solutions are topologically stable in even spacetime dimensions only,
in the Skyrme case they are stable in all dimensions, like in the monopole case.

\bigskip

Concerning the explicit solutions described in this paper, we have considered both flat spacetime (Skyrmions) configurations and self-gravitating solutions (Skyrme stars and BHs with Skyrme hair). Overall, we have unveiled a rich and involved space of solutions.

In the spherically symmetric case,
the pattern of the $d=4$ solutions is recovered,
with a branch of gravitating  Skyrmions emerging from the flat space/Schwarzschild 
background solutions. 
A secondary branch of solutions is also found,
which, however, possesses a different limiting behaviour than in the $d=4$ case.

On the critical behaviour of spherically symmetric solutions as a function of $\alpha$ or $r_H$,
we remark that
some features resemble the case of a higher dimensional gravitating
non-abelian system with higher derivatives terms in addition to
 the usual $F(2)^2$ one \cite{Brihaye:2002jg}.
The clarification of the critical behaviour therein 
has required a reformulation of the problem 
with techniques bases on   a fixed point
analysis of nonlinear ODEs \cite{Breitenlohner:2005hx}.
We expect that a similar approach would 
help to
clarify the critical behaviour 
of the gravitating solutions in this work.

The spinning hairy BH we have reported in this paper are the \textit{first example} of a spinning BH with Skyrme hair, since the corresponding $d=3+1$ solutions have not yet been constructed. One salient feature of these rotating BHs with Skyrme hair is that they possess a static limit. Whereas this is expected, since static BHs with Skyrme hair are known in $d=3+1$ spacetime dimensions, it contrasts in a qualitative way with the behaviour of other BHs with scalar hair, namely Kerr BHs with scalar hair~\cite{Herdeiro:2014goa,Herdeiro:2015gia,Herdeiro:2015tia} or MP BHs with scalar hair~\cite{Brihaye:2014nba,Herdeiro:2015kha}. It is therefore of some interest to expand on the comparison between these two models, since they are both examples of BHs with scalar hair:

\begin{description}
\item[$\bullet$] Skyrmions on Minkowski spacetime are \textit{topological solitons}, in view of their asymptotic boundary conditions. $Q$-balls~\cite{Coleman:1985ki}, on the other hand, which arise in models of self-interacting complex scalars fields, 
but with standard kinetic terms, 
are perhaps the simplest example of a \textit{non-topological} soliton. For the latter, the complex nature of the scalar field is crucial to satisfy Derrick's theorem, allowing for an underlying harmonic time dependence of the scalar field but that vanishes at the level of all physical quantities, thus yielding static or stationary lumps of energy.
\item[$\bullet$] Flat spacetime spinning Skyrmions have a static limit; thus they carry an \textit{arbitrarily small} angular momentum (for given topological charge). Spinning $Q$-balls, on the other hand, have their angular momentum quantised in terms of 
their charge~\cite{Volkov:2002aj,Kleihaus:2005me}, which in this case is a Noether charge, due to a $U(1)$ global symmetry. Thus, they have a \textit{minimum} angular momentum, for given Noether charge, and the spinning solutions are not continuously connected to the static ones. 
\item[$\bullet$] When minimally coupled to Einstein's gravity, self-gravitating Skyrmions become Skyrme stars. But the structure of the model to obtain these solutions \textit{remains the same}, namely the key higher order kinetic term is still mandatory. When minimally coupled to Einstein's gravity, back reacting $Q$-balls become boson stars. But \textit{gravity can replace part of the key structure} of the flat spacetime model: one can get rid of the self-interactions potential and keep only a mass term~\cite{Kaup:1968zz,Ruffini:1969qy}, as the non-linearities of Einstein's gravity are sufficient to counter balance the dispersive nature of the scalar field and create equilibrium boson stars. 
\item[$\bullet$] Likewise, Skyrmions, Skyrme stars can rotate slowly and \textit{connect to the static limit}, whereas rotating boson stars form an infinite discrete set of families \textit{disconnected from static boson stars}~\cite{Yoshida:1997qf,Kleihaus:2005me}, for any model, with or without self-interactions, in any spacetime dimension.
\item[$\bullet$] Static Skyrme stars \textit{admit placing a BH horizon} at their centre, yielding static BHs with Skyrme hair, both in $d=4$~\cite{Luckock:1986tr,Luckock} and $d=5$ (and likely in other dimensions). Static boson stars \textit{do not admit} placing a BH horizon at their centre, as shown by the no-hair theorem in~\cite{Pena:1997cy}.
\end{description}

In spite of all these differences, the spinning BHs with Skyrme hair that we have found in this paper rely on \textit{precisely the same condition} that the Kerr BHs with scalar hair or MP BHs with scalar hair, the latter being the hairy BH generalisation of spinning boson stars in $d=4$ and $d=5$. This condition is the the synchronisation of the phase angular velocity of the Skyrme field and the angular velocity of the horizon, eq.~\eqref{synch}. This is yet another example for the universality of this condition in obtaining spinning hairy BH solutions.

\bigskip

Finally, let us remark that possible avenues for future research include:
$i)$ the investigation of stability of the considered configurations
(based on the analogy with the $d=4$ case, we expect some of the gravitating solutions to be stable);
$ii)$ the construction of less symmetric Skyrmions; the simplest case would be the (higher winding number) axially symmetric solutions and
configurations with $J_1 \neq J_2$. Solutions with discrete symmetry only are also likely to exist in this model;
$iii)$ the construction of $d>5$ generalizations: in Appendix B we present a general framework in this direction.


 \vspace*{0.7cm}
\noindent{\textbf{~~~Acknowledgements.--~}}  
C.H and E.R. gratefully acknowledge support from the FCT-IF programme.  
This work was also partially supported 
by  the  H2020-MSCA-RISE-2015 Grant No.  StronGrHEP-690904, the H2020-MSCA-RISE-2016 Grant No.  FunFiCO-777740, 
and by the CIDMA project UID/MAT/04106/2013.

 \appendix
\section{Solutions with a quartic term}
\setcounter{equation}{0}
\renewcommand{\theequation}{A.\arabic{equation}}

The reported numerical results have $\lambda_2=0$. 
To test their generality,
we have also studied
how the inclusion of a quartic term
affects the properties of the solutions, in a number of cases.
  
Starting with the solutions in flat spacetime background,
we display in Figure 12 the mass $M$ as a function 
of $\lambda_2$ 
for Skyrmions with several values of the frequency $\om$
(note that all solutions displayed in this Appendix are 
for  
$\lambda_0=\lambda_1=\lambda_3=1$).
As expected, one can see that the presence of 
quartic term in the Skyrme action
increases the mass of the solutions, and for $\om\neq 0$, the same trend applies for the angular momentum.
Observe that
the $M(\lambda_2)$-function is almost linear for small frequencies.
It is also interesting to mention
that the $\om=0$ Skyrmions with large enough
values of $\lambda_2$
are well approximated by the self-dual solution
(\ref{ex1}),
the  contribution of the quartic term
dominating the system in this case.

 {\small \hspace*{3.cm}{\it  } }
\begin{figure}[t!]
\hbox to\linewidth{\hss%
	\resizebox{8cm}{6cm}{\includegraphics{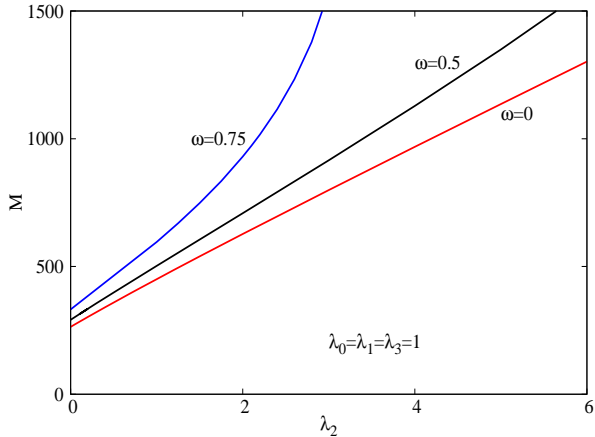}}
\hss}
\caption{\small 
 The  mass $M$ is shown
 as a functions of the parameter $\lambda_2$, for static and  
spinning Skyrmions, with different frequencies.}
\end{figure}

\begin{figure}[ht]
\hbox to\linewidth{\hss%
	\resizebox{8cm}{6cm}{\includegraphics{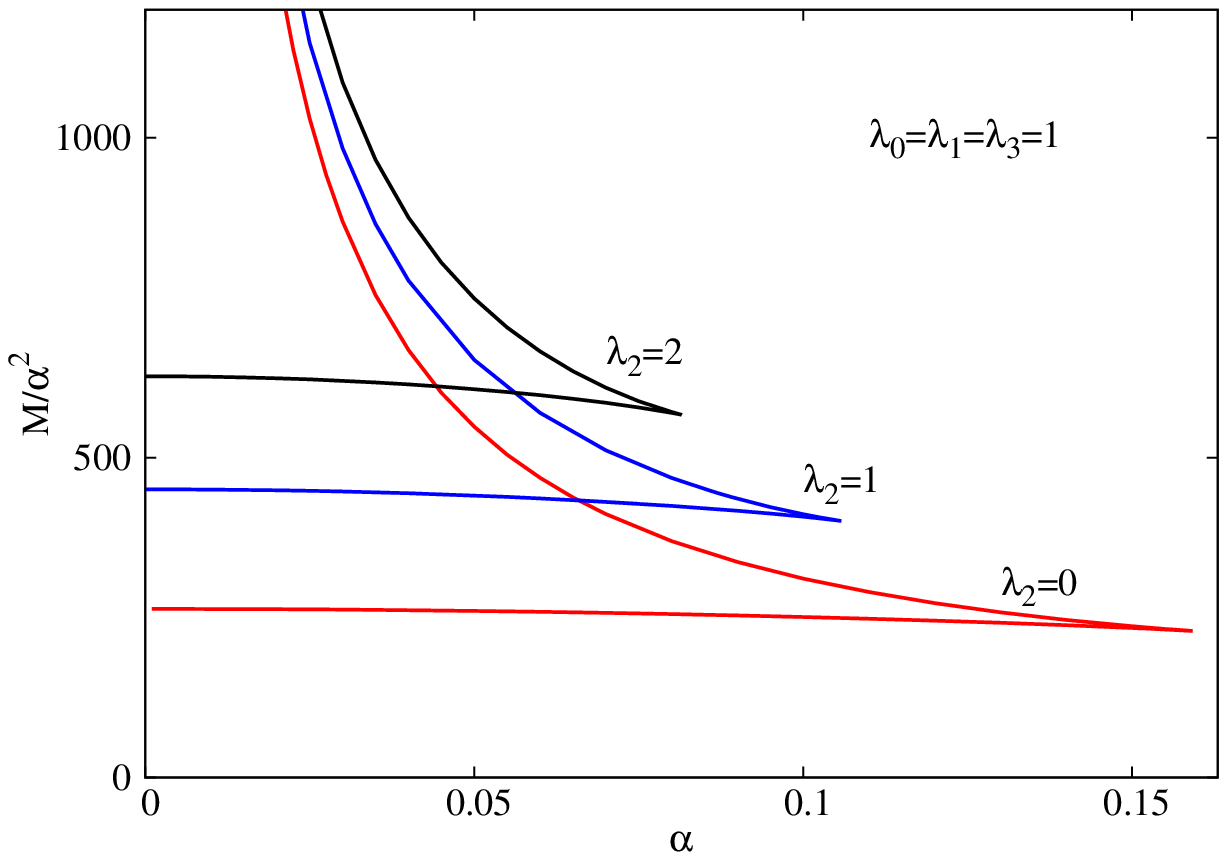}}
\hspace{10mm}%
        \resizebox{8cm}{6cm}{\includegraphics{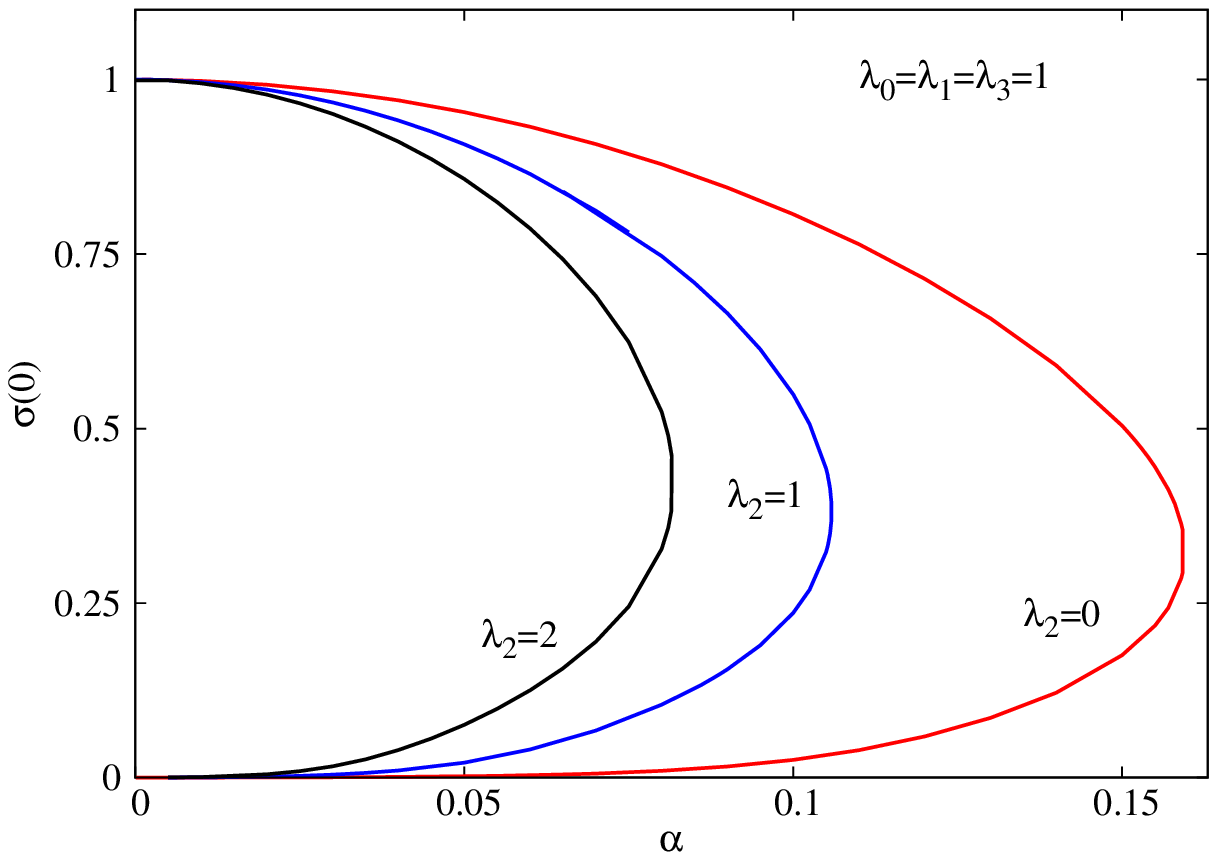}}	
\hss}
\caption{\small 
 The  mass $M$ (left panel) and the parameter $\sigma(0)$ (right panel) are shown as 
a function of the
coupling constant $\alpha$, 
 for Skyrme stars in our model with different values of $\lambda_2$.
}
 \end{figure}
\begin{figure}[ht]
\hbox to\linewidth{\hss%
	\resizebox{8cm}{6cm}{\includegraphics{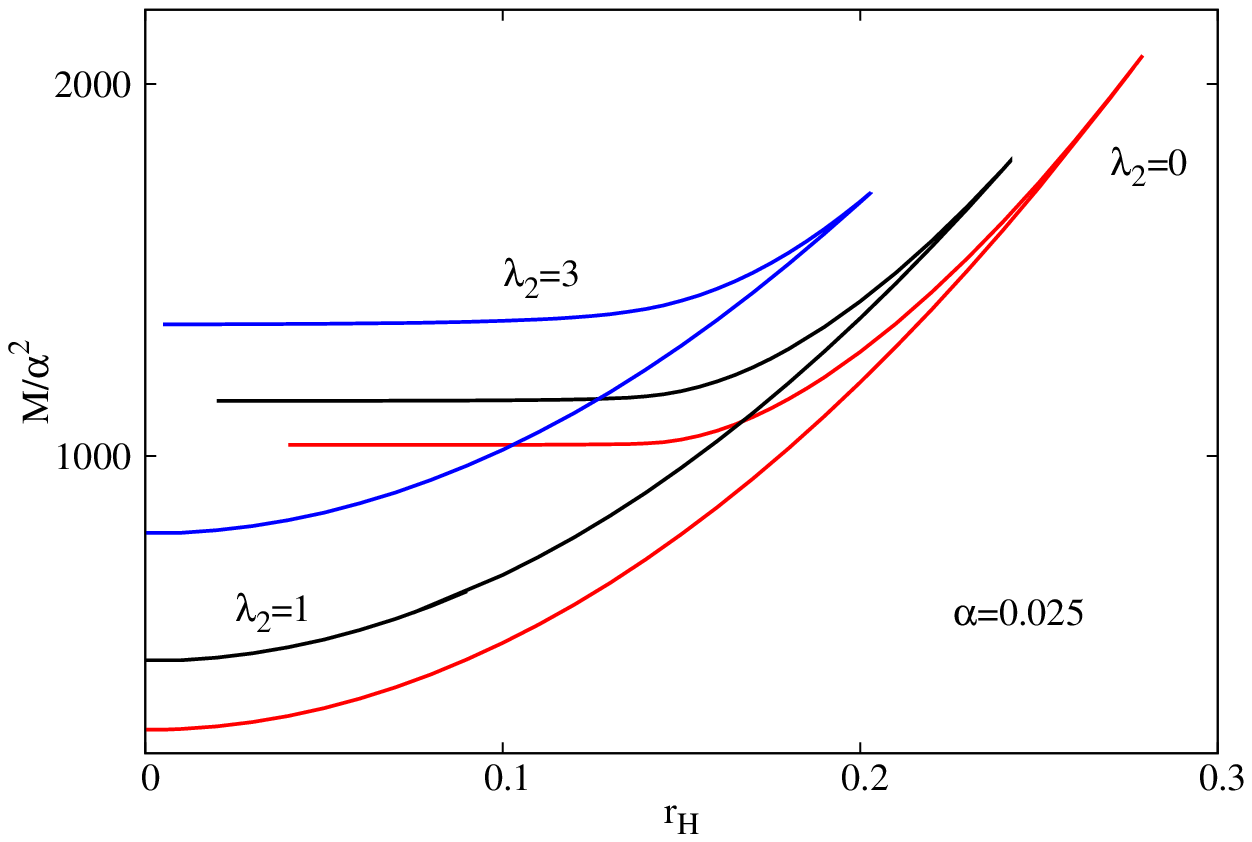}}
\hspace{10mm}%
        \resizebox{8cm}{6cm}{\includegraphics{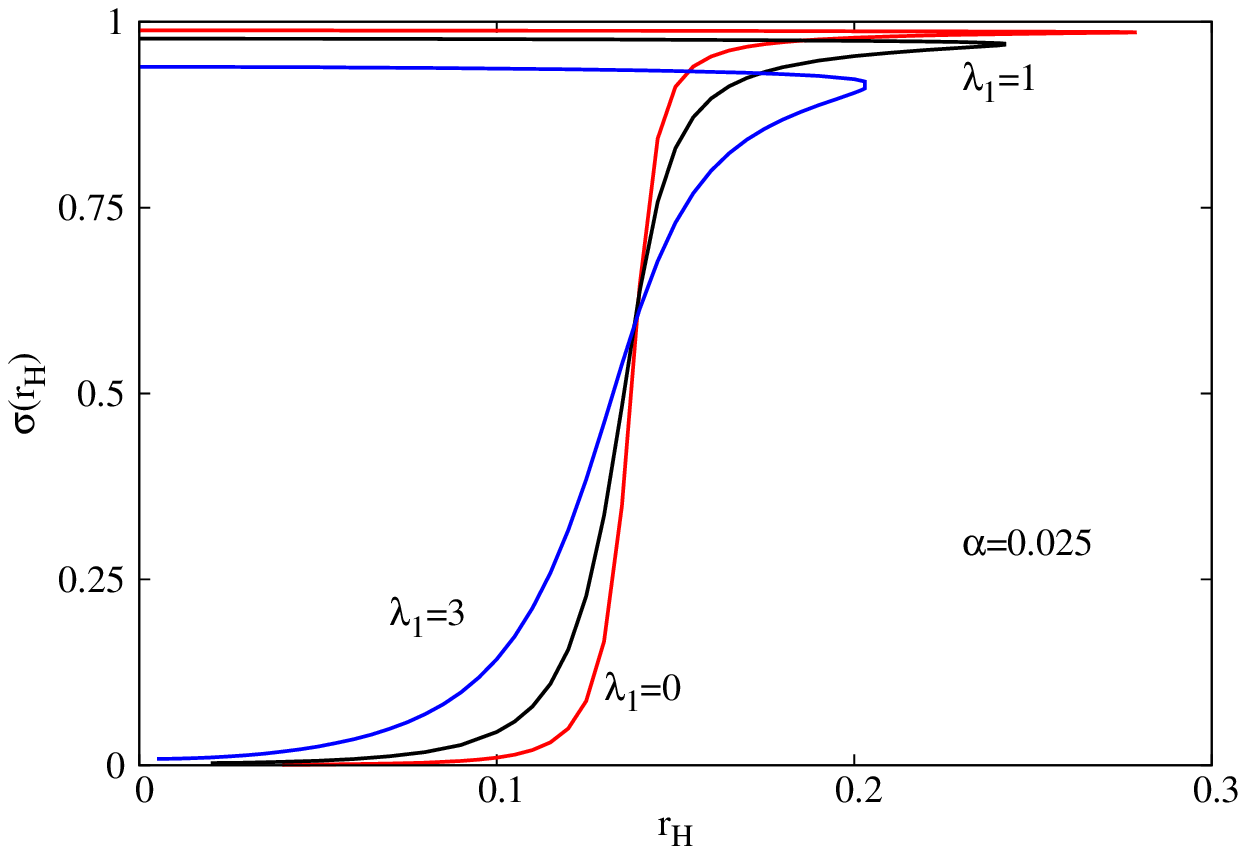}}	
\hss}
\caption{\small 
 The  mass $M$ (left panel) and the parameter $\sigma(r_H)$ (right panel) are shown for spherical BHs with Skyrme hair, as 
a function of the horizon radius  
 for several values of $\lambda_2$ and a fixed $\alpha$.
}
 \end{figure}

Turning now to gravitating solutions, we display in Figure 13
families of spherically symmetric Skyrme stars with
several values of $\lambda_2$.
One can see that the picture found in Section 4.1 appears to be generic,
with the occurrence of two branches of solutions in terms of $\alpha$.
Also, one notices the existence of a maximal value of $\alpha$, which decreases with increasing $\lambda_2$,
while the limiting behaviour on the second branch is similar to that found for 
solutions without a quartic term.

Finally, the same conclusion is reached in the presence of an horizon,
$cf.$ Figure 14,
where we show the $(r_H,M)$ and $(r_H,\sigma(r_H))$
diagrams for spherically symmetric BHs with Skyrme hair
with a fixed value of $\alpha$
and three values of $\lambda_2$.

To summarize, the presence of a quartic term
in the Skyrme action does not seem to lead to new qualitative features,
at least for the range of parameters considered.

\section{$O(D+1)$ Skyrme models on a $\R^{D}$-Euclidean space}
\setcounter{equation}{0}
\renewcommand{\theequation}{A.\arabic{equation}}

The Skyrme model~\cite{Skyrme:1961vq,Skyrme:1962vh} in $3+1$ ($i.e.$, $D=3$) dimensions\footnote{In this Section, we shall take $d=D+1$, $i.e.$ $D$ denotes the number of space dimensions.}
is a nonlinear chiral field theory which supports topologically stable solitons in the static limit. These solitons, which are called Skyrmions, describe baryons and nuclei.
In its original formulation~\cite{Skyrme:1961vq,Skyrme:1962vh}, 
the model is described by an $SU(2)$ valued field $U$. The Skyrmions are stabilised by a topological charge which is characterised
by the homotopy class $\pi_3(SU(2))=\mathbb{Z}$.

Alternatively, the chiral matrix $U$ can be parametrised as $U=\f^a\si_a$ ($a=1,\dots,4$) and its inverse as $U^{-1}=U^{\dagger}=\f^a\tilde\si_a$, where $\si^a$ and $\tilde\si_a$ are the chiral representations of the algebra of $SU(2)$.
The scalar $\f^a$ is subject to the constraint $\phi^a\phi^a = 1$, such that it takes its values on $S^3$, the latter being
parametrised by the angles parametrising the element $U$ of the group $SU(2)$. The homotopy class in terms of $\f^a$ is now $\pi_3(S^3)=\mathbb{Z}$. This is the parametrisation that will be adopted here.
 
The parametrisation of the Skyrme scalar in terms of the chiral field $U$ is peculiar to $D=3$.
Indeed in $D=2$ the famous Belavin-Polyakov vortices~\cite{Polyakov:1975yp} of the $O(3)$ sigma model on $\R^2$ are parametrised by the $S^2$ valued scalar~\footnote{Indeed in this low dimensional case too, there is an alternative
parametrisation of the ``Skyrme scalar'', namely in terms of the $\mathbb C\mathbb P^1$ field $z^\al$ ($\al=1,2$) subject to $z^\dagger z=1$, by virtue of the equivalence of the $O(3)$ and $\mathbb C\mathbb P^1$
sigma models $via$ $\f^a=z^\dagger\si^az$.} 
subject to $\f^a\f^a = 1$ ($a=1,2,3$), pertaining to the homotopy class $\pi_2(S^2)=\mathbb{Z}$.

In all dimensions $D\geqslant 4$, these low dimensional accidents are absent so Skyrme models are defined as
the $O(D+1)$ sigma models on $\R^{D}$, described by the Skyrme scalar $\f^a$, $a=1,2,\dots,D+1$ subject to the constraint
$\phi^a \phi^a=1$.
\be
\label{constr}
|\f^a|^2=1\ ,\quad a=1,2,\dots,D+1\,,
\ee
pertaining to the homotopy class $\pi_D(S^D)=\mathbb{Z}$.

In any given dimension $D$, the energy density functional ${\cal H}^{(D)}$ can be endowed with a ``potential'' term, $e.g.$ the
``pion mass'' type potential
\be
\label{potD}
V=1-\f^{D+1}~,
\ee
and $D$ possible ``kinetic'' terms ${\cal H}^{(p,D)}$, which are defined as follows.

Employing the shorthand notation for the $1$-form
\be
\label{notation}
\f(1)=\f_{i}^{a}\stackrel{\rm def}=\pa_{i}\f^{a}\ ,\quad i=1,2,\dots D\ ;\quad a=1,2,\dots,D+1\,,
\ee
one defines the $p$-form
\be
\label{pform}
\f(p)=\f_{i_1i_2\dots i_p}^{a_1a_2\dots a_{p}}\stackrel{\rm def}=\f_{[i_1}^{a_1}\f_{i_2}^{a_2}\dots\f_{i_p]}^{a_p}~,
\ee
which is the $p$-fold product of $\f(1)=\f_i^a$, totally anitisymmetrised in the indices $i_1,i_2,\dots, i_p$.

In this notation, the kinetic terms ${\cal H}^{(p,D)}$ are concisely defined as
\bea
\label{Hp}
{\cal H}^{(p,D)}&=&|\f(p)|^2
\eea
such that only the square of any 'velocity' field $\f_i^a$ occurs in ${\cal H}^{p,(D)}$.

In this notation, the most general energy density functional in any dimension $D$ is
\be
\label{HD}
{\cal H}^{(D)}=\sum_{p=1}^{p=D}\la_p{\cal H}^{(p,D)}+\la_0V\,,
\ee
and the topological charge density (up to normalisation) is
\bea
\label{topch}
\vr^{(D)}&\simeq&\vep_{i_1i_2\dots i_D}\vep^{a_1a_2\dots a_{D+1}}\f_{i_1}^{a_1}\f_{i_2}^{a_2}\dots\f_{i_D}^{a_D}\f^{a_{D+1}}\\
&=&\vep_{i_1i_2\dots i_D}\vep^{a_1a_2\dots a_{D+1}}\pa_{i_1}\f^{a_1}\pa_{i_2}\f^{a_2}\dots\pa_{i_D}\f^{a_D}\f^{a_{D+1}}\,.\nonumber
\eea
It is well known that \re{topch} is $essentially\ total\ divergence$ 
and hence qualifies as a topological charge density. To see this, subject the quantity
\[
\tilde\vr=\vr^{(D)}+\la(1-|\f^a|^2)
\]
to variations $w.r.t.$ 
the scalar field $\f^a$, taking account of the Lagrange multiplier $\la$. The result is $0=0$, as expected from a density which is $total\ divergence$. Alternatively, one can employ a parametrisation
of $\f^a$ that is compliant with the constraint \re{constr}, $e.g.$ when employing a particular Ansatz. In that case $\vr^{(D)}$ itself would take an $explicitly\ total\ divergence$ form \cite{Tchrakian:2015pka}.

To express the Bogomol'nyi inequalities of this system, we define the Hodge dual of the $(D-p)$-form $\f(D-p)$, which is the $p$-form as
\be
\label{pHodge}
^{\star}\f(p)\stackrel{\rm def}= {^{\star} \f_{i_1\dots i_{p}}^{a_1\dots a_{p}}}=\frac{1}{p!(D-p)!}\vep_{i_1\dots i_pi_{p+1}\dots i_{D}}\vep^{a_1\dots a_pa_{p+1}\dots a_Da_{D+1}}\,\f_{i_{p+1}\dots i_D}^{a_{p+1}\dots a_D}\f^{a_{D+1}}\,.
\ee
For any given $D$, one can now state the $D$ inequalities in terms of \re{pform} and \re{pHodge}, each labelled by $p$, as
\be
\label{ineqpD}
\left|\f(p)\mp\ka^{(D-2p)}\, {^{\star}}\f(p)\right|^2\geqslant 0\ ,\quad p=1,2,\dots D\,,
\ee
in which $\ka$ is a constant with dimension $L^{-1}$ compensating for the difference between the dimensions of $\f(p)$ and ${^{\star}}\f(p)$, which are different in the general case $D\neq 2p$.

It is clear from the definitions \re{pform} and \re{pHodge} that the cross-terms
in  the inequalities \re{ineqpD}, for each $p$, is proportional to the topological charge density \re{topch}.
It follows that the $p$ inequalities \re{ineqpD} lead to the required lower bound on the energy
\be
\label{lbp}
{\cal H}^{(D)}\geqslant \vr^{(D)}\implies\int_{\R^D}{\cal H}^{(D)}\geqslant \int_{\R^D}\vr^{(D)}\,,
\ee
provided of course that the potential \re{potD} is by definition positive definite.

The best known examples of such topologically stable Skyrmions are the Belavin--Polyakov (self-dual) vortices~\cite{Polyakov:1975yp}  on $\R^2$, and the familiar Skyrmions~\cite{Skyrme:1962vh} on $\R^3$. In the $D=2$ case the most general
system \re{HD} is that with all coefficients $\la_0,\la_1,\la_2$ present 
and in the $D=3$ case the most general one is that with all coefficients $\la_0,\la_1,\la_2,\la_3$ present.
Of course, which terms must be retained in each case is
governed by the requirement of Derrick scaling.

Of the $D$ inequalities \re{ineqpD} only one can be solved with power decaying solutions at $r\to\infty$ on $\R^D$, namely the one for which $D=2p$ when the dimensional constant $\ka$ does not appear. Furthermore in this case the model \re{HD} must consist
exclusively of the $p=D/2$ term in the sum, $i.e.$ $\la_p=0$ for $p\neq(D/2)$, otherwise the system will be overdetermined. Such a model has solutions that saturate the topological lower bound \re{lbp}. In all other cases, when $\la_p\neq 0$ for $p\neq(D/2)$,
the solutions cannot saturate the lower bound \re{lbp}, which includes Skyrmions in all odd Euclidean~\footnote{The (anti-)self-duality equations resulting from \re{ineqpD} on $S^D$ for odd $D$, namely
\[
\f(p)\mp\ka^{(D-2p)}\, {^{\star}}\f(p)=0
\]
can be solved, since in that case the dimensional constant $\ka$ is absorbed by the radius of the sphere. See $e.g.$~\cite{Manton:1986pz,Manton:1987xt,Canfora:2014aia}.
} dimensions $\R^D$.

In particular in the important case of $\R^3$, there exist no solutions to first-order (anti-)self-duality equations saturating the lower bound. Exact solutions
to the second-order equations can be constructed only numerically and not in closed form. However,
approximate solutions on $\R^3$ in closed form are known. For example the rational map ansatz~\cite{Houghton:1997kg}, and the Atiyah-Manton~\cite{Atiyah:1989dq,Manton:1990gr} construction where the holonomy of the Yang-Mills instantons
on $\R^4$ gives a good approximation for the Skyrmion on $\R^3$. This last approach is extended to give an approximate construction for the Skyrmion on $\R^7$ by exploiting the holonomy of the Yang-Mills instantons
on $\R^8$ in \cite{Nakamula:2016wwv}, which possible be extended to higher dimensions $modulo\ 4$.

The first-order (anti-)self-duality equations for the $p$-Skyrme system ${\cal H}^{(p,D)}=|\f(p)|^2$ on $\R^{2p}$
\be
\label{asd}
\f(p)=\pm\, {^{\star}}\f(p)
\ee
can be solved in closed for subject to no symmetries only in the case $D=2$. In that case~\cite{Polyakov:1975yp} the equations \re{asd} reduce to the Cauchy-Riemann equations. In all higher dimensions $D=2p>2$, only solutions of the
system subject to radial symmetry are known, and the form factors $F(r)$ for all $(p,D=2p)$ are given by the same function \re{ex1}. It is interesting to remark here that this situation hold also in the case of the $p$ hierarchy of BPST
instantons~\cite{Tchrakian:1984gq,Grossman:1984pi} on $\R^{4p}$.

It is interesting to push the analogy between the self-duality equations of the 
$p$-Yang-Mills systems on $\R^{4p}$
and the $p$-Skyrme systems on $\R^{2p}$. In both cases the spherically symmetric solutions are described by the same radial
function in all dimensions. In both cases, these equations become more overdetermined with increasing dimension. In the case of the $p$-YM equations, axially symmetric solutions (where spherical symmetry was imposed in the $\R^{4p-1}$ subspace of
$\R^{4p}$) were found~\cite{Chakrabarti:1985qj}, but imposing less stringent symmetry rendered the self-duality equations overdetermined~\cite{Tchrakian:1990gc}. It turns out that a similar situation holds for the first-order $p$-Skyrme
equations. In this case the only solutions in $D=2p\ge 4$ known are the radially symmetric ones, with the axially symmetric equations (when spherical symmetry is imposed in the $\R^{2p-1}$ subspace of $\R^{2p}$) turn out to be overdetermined
(See Appendix of Ref,~\cite{Tchrakian:1990gc}).

As a final remark, one notes that the first-order self-duality equations of the $p$-Yang-Mills system coincide with the first-order self-duality equations of the $p$-Skyrme equation \re{ex12} with the replacement $w(r)=\cos F(r)$.
But instanton and instanton-antiinstanton solutions in $\R^4$ are known in the case of $p=1$ Yang-Mills
\cite{Radu:2006gg}.
 This raises the question whether Skyrmion--anti-Skyrmion
solutions may also exist for the $p$-Skyrme model on $\R^{2p}$?

 \begin{small}

 \end{small}

\end{document}